\pdfoutput=1
\documentclass[cits,hyper]{JINST}

\usepackage[squaren]{SIunits}

\usepackage{amsmath}
\newcommand{\di}{\ensuremath{\,\text{d}}}         
\newcommand{\Poi}{\ensuremath{\text{Poi}}}        
\newcommand{\Ga}{\ensuremath{\text{Ga}}}          

\newcommand{\doi}[1]{\href{http://dx.doi.org/#1}{doi: #1}}
\newcommand{\arxiv}[1]{\href{http://arXiv.org/abs/#1}{arXiv: #1}}

\title{Reference analysis of the signal + background model in counting
  experiments II.  Approximate reference prior}

\author{Diego Casadei\\
 \normalsize \emph{FHNW, School of Engineering, Bahnhofstrasse 6, 5210
   Windisch, Switzerland, and
   \\
   UCL, Department of Physics and Astronomy, Gower Street, London
   WC1 E 6BT, UK}  \\
 \email{diego.casadei@cern.ch}
}

\abstract{
  The objective Bayesian treatment of a model representing two
  independent Poisson processes, labelled as ``signal'' and
  ``background'' and both contributing additively to the total number
  of counted events, is considered.  It is shown that the reference
  prior for the parameter of interest (the signal intensity) can be well
  approximated by the widely (ab)used flat prior only when the
  expected background is very high.
  On the other hand, a very simple approximation (the limiting form of
  the reference prior for perfect prior background knowledge) can be
  safely used over a large portion of the background parameters space.
  The resulting approximate reference posterior is a Gamma density
  whose parameters are related to the observed counts.  This limiting
  form is simpler than the result obtained with a flat prior, with the
  additional advantage of representing a much closer approximation to
  the reference posterior in all cases.  Hence such limiting prior
  should be considered a better default or conventional prior than the
  uniform prior.
  On the computing side, it is shown that a 2-parameter fitting
  function is able to reproduce extremely well the reference prior for
  any background prior.  Thus, it can be useful in applications
  requiring the evaluation of the reference prior for a very large
  number of times.
\\
  \emph{The published version JINST 9 (2014) T10006
   (\doi{10.1088/1748-0221/9/10/T10006}) has a typo in the
   normalization constant $N$ of eq.~(\ref{eq-lim-ref-posterior}),
   fixed here.}
}


\begin{document}

\maketitle


 \section{Introduction}

 This document complements and extends the results shown in
 Ref.~\cite{casadei2012} (hereafter Paper I), in which the reference
 analysis is performed of the signal + background model in counting
 experiments, when partial information is available about the
 background and an objective Bayesian solution is desired.  In the
 model, signal and background events come from two independent Poisson
 sources, so that the total number $n\ge0$ of observed counts is
 distributed accordingly to
 \begin{equation}\label{eq-poisson}
   \Poi(n|s+b) = \frac{(s+b)^n}{n!} \, e^{-(s+b)}
 \end{equation}

 The goal is to perform statistical inference on the signal strength
 $s\ge0$, hence the background strength $b\ge0$ is a nuisance
 parameter.  The starting point is the Bayes' theorem
 \begin{equation}\label{eq-bayes-theorem}
   p(s,b|n) \propto \Poi(n|s+b) \, p(s) \, p(b)
 \end{equation}
 which gives the joint posterior probability density $p(s,b|n)$ of
 signal and background strengths, given the observed number $n$ of
 events.  The joint posterior is proportional to the product of the
 likelihood function --- that is (\ref{eq-poisson}) when considered as
 a function of $s$ and $b$ for fixed $n$ --- with the prior densities
 $p(s) \, p(b)$ of signal and background.  After integrating over $b$,
 one gets the marginal posterior density $p(s|n)$ which represents the
 full solution of the inference problem.  From $p(s|n)$ one can
 compute summary information like e.g.~the posterior expectation or
 most probable value, enclosed by intervals representing some given
 probability, say 68.3\% or 95\% posterior probability.

 In Paper I the very common situation is considered in which one
 claims no prior information on the signal $s$, but does have prior
 estimates of the background expectation $E[b]$ and standard deviation
 $\sqrt{V[b]}$ (the square root of the prior variance).  These two
 values are sufficient to specify uniquely the prior density, if the
 latter is chosen to be a Gamma density
 \begin{equation}\label{eq-bkg-prior}
   p(b) = \Ga(b|\alpha,\beta)
        = \frac{\beta^\alpha}{\Gamma(\alpha)}
           \, b^{\alpha-1} \, e^{-\beta b}
 \end{equation}
 (the conjugate prior of the Poisson model) with shape parameter
 $\alpha>0$ and rate parameter $\beta>0$ fixed by requiring $E[b] =
 \alpha/\beta$ and $V[b] = \alpha/\beta^2$.

 The same Poisson model (\ref{eq-poisson}), when assuming no prior
 knowledge about \emph{both} the signal and the background, has been
 addressed by \cite{demortier2010} and \cite{pierini2011}, where a
 reference prior is found for both signal and background.  However in
 practical applications, expecially when a search for rare or new
 phenomena is performed, the background is known quite well (the
 discovery of the Higgs boson is one example
 \cite{higgs-atlas,higgs-cms}).  Hence we restrict ourselves to the
 inference problems about the signal strength $s$, when there is at
 least some knowledge of the background yield summarized by the prior
 background expectation $E[b]$ and variance $V[b]$ (or standard
 deviation).

 Paper I finds the reference prior for the signal starting from the
 marginal model
 \begin{equation}\label{eq-marginal-model}
   P(n|s) = \int_0^\infty \Poi(n|s+b) \,
                         \Ga(b|\alpha,\beta) \di b
 \end{equation}
 and following the algorithm explained in \cite{Sun1998}, which
 requires the computation of the Fisher's information function
 \begin{equation}\label{eq-fisher-info}
   I(s) = - E\left[ \left( \frac{\partial^2}{\partial s^2}
            \log p(k|s) \right) \right] 
 \end{equation}
 As the resulting reference prior $\pi(s) \propto |I(s)|^{1/2}$ is not
 integrable over the positive real line, it has an arbitrary scale
 factor.  It should be emphasized that, provided that the
 corresponding posterior is a proper density, this is not a problem
 for the reference prior, which is formally constructed in such a way
 to maximize the amount of missing prior information and does not
 represent one's degree of belief
 \cite{BayesianTheory1994,Bernardo2005a,Bernardo2009a}.  The choice
 made in Paper I is to define
 \begin{equation}
   \label{eq-ref-prior}
   \pi(s) = \frac{|I(s)|^{1/2} }{|I(0)|^{1/2} }
 \end{equation}
 which makes it trivial to compare it against the uniform
 prior, so widespread that it can be considered a conventional prior.

 As noted in Paper I (and references therein), the flat prior is often
 presented as ``noninformative'', which is not the case, as for the
 current problem it is not the result of a formal procedure to
 construct an objective prior\footnote{For discrete distributions, the
   flat prior maximizes the entropy and coincides with the reference
   prior, but this is not generally true in the continuous case, nor
   for the model considered here.  When the reference prior for a
   particular parametrization of a continuous model is flat, such
   parametrization is called ``reference parametrization''
   \cite{Bernardo2005a}.}.  At the same time, it can not be an
 informative prior, because it is not normalizable (hence can not
 represent a degree of belief): strictly speaking it has no formal
 justification.  In facts, the (ab)use of the uniform prior is the
 most important source of criticism toward the Bayesian approach in
 scientific inference.  Nevertheless, the flat prior \emph{may}
 provide a good approximation to the reference prior in some case,
 although this is not a general rule.

 The function (\ref{eq-ref-prior}) is monotonically decreasing and
 depends on the shape and rate parameters describing the prior of the
 background.  It is flatter for increasing background expectation
 $E[b]=\alpha/\beta$ and for decreasing prior background relative
 uncertainty $\sigma[b]/E[b] = 1/\sqrt{\alpha}$.
 Thus for increasing $\alpha$ and decreasing $\beta$, the flat prior
 may provide a good approximation to an objective prior, although it
 is the user's responsibility to decide when the difference with
 respect to $\pi(s)$ is acceptably small.

 The reference prior $\pi(s)$ obtained in Paper I has the form of an
 infinite series, as shown in the next section.  Although its
 implementation in a computer program has good performance when
 following the recommendations of Appendix A of Paper I, some users
 consider it too complicated and error-prone to write the corresponding
 code.  For this reason, they end up into using the flat prior as the
 ``conventional'' choice, despite from the known issues of this prior
 (it is mathematically ill-defined and many people do not consider it
 ``objective'', as it gives absurdly high weights to large values of
 the signal, which are known not to be true).  Hence it is important
 to find the cases in which the use of $\pi(s)$ is \emph{not}
 required.

 Luckily enough, a very simple expression exists for the limiting case
 of perfect background knowledge, which provides a good approximation
 in many practical problems and is closer to the reference prior than
 the flat prior.  The corresponding posterior is a Gamma function
 which approximates the reference posterior much better than the
 posterior obtained with a flat prior (the marginal model given by
 eq.~(\ref{eq-marginal-model}) below).  This solution is so simple
 that there is really no motivation to adopt a flat prior.  Below, we
 will see how well this approximation works.\footnote{A movie
   available on
   \href{https://www.youtube.com/watch?v=vqUnRrwinHc}{https://www.youtube.com/watch?v=vqUnRrwinHc}
   shows $\pi(s)$ for a wide range of parameter values, comparing it
   to the uniform prior and to the approximate reference priors
   illustrated below.}
 Even when it is not a perfect approximation, it is usually closer to
 the full reference prior than the constant prior.  For this reason,
 it is recommended as the default or conventional prior whenever the
 use of the full reference prior is considered too complicated.

 In practical applications, the performance can be improved if the
 reference prior is evaluated only at few tens of points and
 interpolated with a fitting function to be used in the following
 computations, as a probability density function may need to be
 evaluated a very large number of times.
 The limiting properties of the reference prior suggest the form of a
 1-parameter fitting function which is easy to program and provides a
 good approximation to the reference prior over a wide portion of the
 parameters space.  Furthermore, a 2-parameter approximate reference
 prior, available in closed form and very quick to compute, is
 practically equivalent to $\pi(s)$ for the entire parameters range
 scanned in this work.  This fitting function can be used to speed-up
 the computation of the reference prior with no precision losses.
 Thus it is has been exploited in the \emph{Bayesian Analysis Toolkit}
 \cite{BAT2009}, which is the first publicly available implementation
 of the reference prior $\pi(s)$ known to the author.
 Appendix~\ref{sec-bat} shows an example of C++ code to be used with
 BAT.  For the users who do not rely on BAT, a table of values of
 $\pi(s)$ for $s\in[0,70]$ for all background parameters examined in
 section~\ref{sec-f0} is freely available on Zenodo
 (\doi{10.5281/zenodo.11896}), to obtain quick approximations as
 described below and in Appendix~\ref{app-data}.


 \section{The reference prior and posterior densities}

 Although the marginal model (\ref{eq-marginal-model}) does not depend
 explicitly on the background, it still depends on the background
 shape $\alpha$ and rate $\beta$ parameters via the integration.
 Paper I shows that the marginal model can be written as
 \begin{equation}\label{eq-marg-mod-2}
   P(n|s) = \left( \frac{\beta}{1+\beta} \right)^{\!\alpha}
           e^{-s} \, f(s;n,\alpha,\beta)
 \end{equation}
 where the polynomial
 \begin{equation}\label{eq-f}
  f(s;n,\alpha,\beta) = \sum_{m=0}^{n} \binom{\alpha+m-1}{m}
                        \frac{s^{n-m}}{(n-m)! \, (1+\beta)^{m}}
 \end{equation}
 behaves like $(s+\alpha/\beta)^n$ when both $\alpha,\beta$ are very large.


 Hence the reference prior also depends on the background parameters,
 and it does so via the Fisher's information function
 \begin{equation}\label{eq-sqrt-fisher-info}
   |I(s)|^{1/2}  =  \left|
       \left( \frac{\beta}{1+\beta} \right)^{\!\alpha} e^{-s} \,
           \sum_{n=0}^{\infty}
           \frac{[f(s;n,\alpha,\beta)]^2}{f(s;n+1,\alpha,\beta)}
       - 1
                 \right|^{1/2}
 \end{equation}
 which involves an infinite sum over terms featuring the polynomial
 (\ref{eq-f}).  This function is not integrable, hence one is free to
 choose a multiplicative constant.  We define the reference prior as
 in eq.~(\ref{eq-ref-prior}), which is the recommended expression in
 practical computations.

 The marginal reference posterior for the signal yield $s$ is 
 \begin{equation}\label{eq-ref-posterior}
   p(s|n) \propto \left( \frac{\beta}{1+\beta} \right)^{\!\alpha}
           e^{-s} \, f(s;n,\alpha,\beta) \, \pi(s)
 \end{equation}
 which is always a proper density, hence the normalization constant
 is just the integral of the expression above.\footnote{In practical
   applications, the constant $[\beta/(1+\beta)]^\alpha$ can be
   dropped from eq.~(\ref{eq-ref-posterior}), before computing the
   normalization constant.  Here we retain it to find the limit of
   certain prior knowledge by letting $\alpha,\beta$ go to infinity.}

 In Paper I it was shown that, in the limiting case of perfect prior
 information about the background,
 the reference prior becomes proportional to $(s+b_{0})^{-1/2}$.  This
 is the Jeffreys' prior for the variable $s' \equiv s+b_{0}$, a quite
 natural result.
 More precisely, the limiting prior which matches the convention that
 $\pi(0)=1$ is
 \begin{equation}\label{eq-lim-ref-prior}
   \pi_0(s) \equiv \sqrt{\frac{b_{0}}{s+b_{0}}}
 \end{equation}
 when one has perfect prior knowledge of the background.
 It is then interesting to check that the posterior also matches the
 result obtained with Jeffreys' prior, which for $n$ observed counts
 is $p(s'|n) = \Ga(s'|n+\frac{1}{2},1)$.

 By substituting $\alpha = b_0 \beta$ in (\ref{eq-f}) and taking the
 limit $\beta\to\infty$ while keeping $\beta=\alpha/b_{0}$ constant,
 such that the background prior (\ref{eq-bkg-prior}) tends to
 $\delta(b-b_0)$, one gets
 \begin{equation*}
   \begin{split}
     f(s;n+1,\alpha,\beta) & \xrightarrow[\beta=\alpha/b_{0}]{\alpha\to\infty}
     \frac{s^n}{n!}  + \frac{s^{n-1} \, b_0}{(n-1)!}
       + \frac{s^{n-2} \, b_0^2}{2(n-2)!} + \frac{s^{n-3} \, b_0^3}{3!(n-3)!} 
       + \cdots + \frac{b_0^n}{n!}
       \\
       & = \sum_{k=0}^n \binom{n}{k} \, s^{n-k} \, b_0^k
         = (s + b_0)^n
   \end{split}
 \end{equation*}
 In the same limit, $[\beta/(1+\beta)]^{\alpha} \to e^{-b_0}$ and
 eq.~(\ref{eq-ref-posterior}) gives \( p_0(s|n) \propto e^{-(s+b_0)}
 \, (s + b_0)^{n-1/2} \) i.e.~the kernel of the Gamma density
 $\Ga(s+b_0|n+\tfrac{1}{2},1)$, as expected.  In conclusion, the
 properly normalized limiting reference posterior is
 \begin{equation}\label{eq-lim-ref-posterior}
   p_0(s|n) = \frac{1}{N} \Ga(s+\tfrac{\alpha}{\beta}|n+\tfrac{1}{2},1)
 \end{equation}
 where the normalization constant is 
 \begin{equation*}
   N = \int_{0}^\infty \Ga(s+\tfrac{\alpha}{\beta}|n+\tfrac{1}{2},1) \di s
     = \int_{\alpha/\beta}^\infty \Ga(x|n+\frac{1}{2},1) \di x
 \end{equation*}

 Compared to the posterior obtained with a flat prior, proportional to
 the marginal model (\ref{eq-marg-mod-2}) (see also
 Appendix~\ref{sec-marg-post}), the limiting reference posterior is
 simpler and easier to implement (most data analysis packages include
 the Gamma function, so no coding is needed).  Hence it is a much more
 appealing conventional solution, provided that it behaves not much
 worse than the flat prior.  Actually it comes out that the limiting
 reference posterior (\ref{eq-lim-ref-posterior}) is a much better
 approximation to the reference posterior than (\ref{eq-marg-mod-2}),
 so that there is really no reason to continue using the flat prior.

 In the following section, we will investigate the performance of the
 approximate reference prior obtained by setting $b_0=E[b]$ in
 eq.~(\ref{eq-lim-ref-prior}).  This is equivalent to ignoring the
 uncertainty on the background in the signal region.  Hence we expect
 it to work well whenever the relative uncertainty $\sqrt{V[b]}/E[b] =
 1/\sqrt{\alpha}$ is small, that is for large values of the shape
 parameter.  The good surprise is that, even when the relative
 uncertainty is sizable, $\pi_0(s)$ is very often a much better
 approximation to the reference prior than a flat prior.


 \section{Properties of the limiting reference prior}\label{sec-f0}

\begin{figure*}[t!]
   \begin{minipage}{0.5\linewidth}
     \centering
     \includegraphics[width=\textwidth]{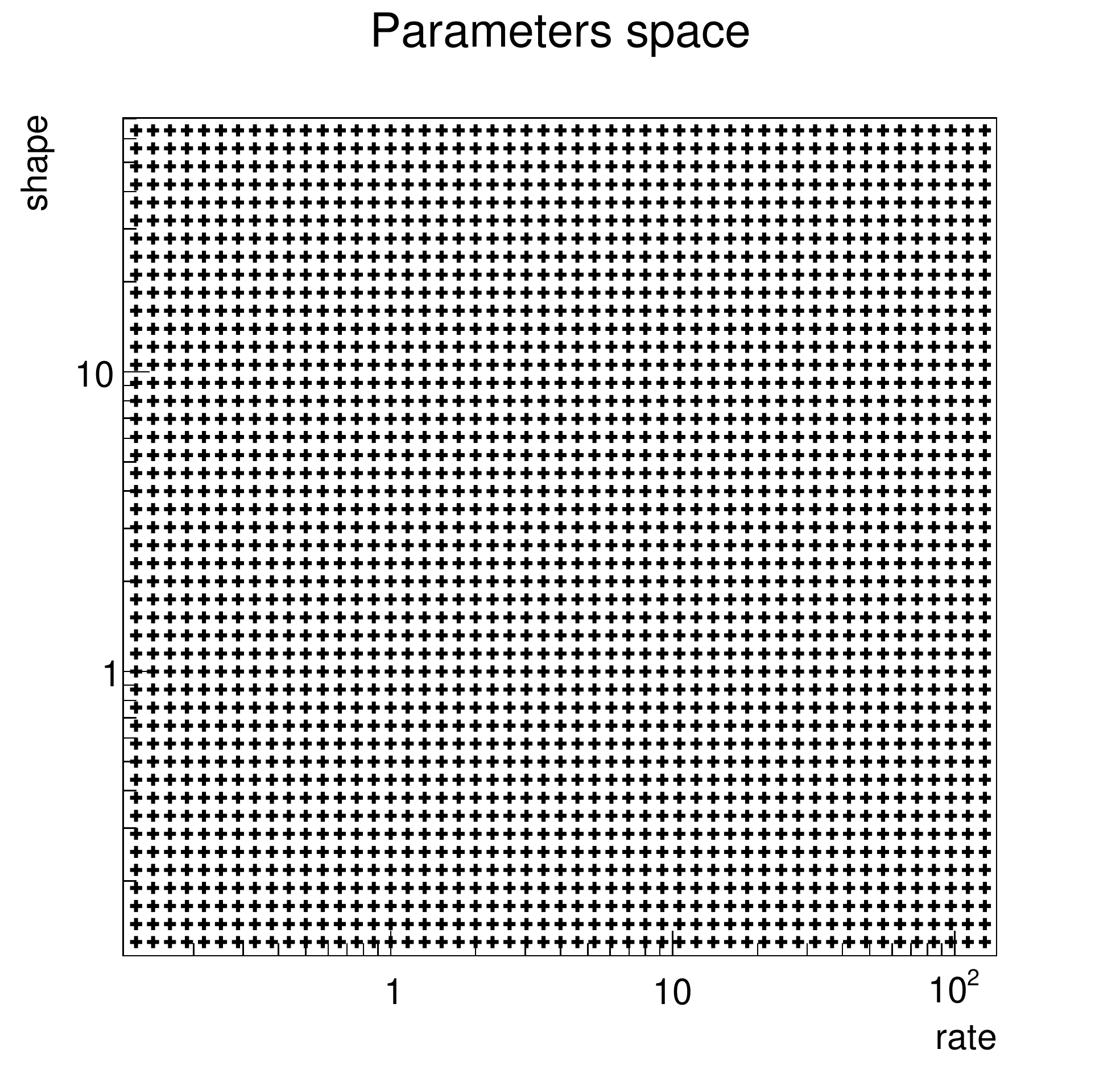}
   \end{minipage}%
   \begin{minipage}{0.5\linewidth}
     \centering
     \includegraphics[width=\textwidth]{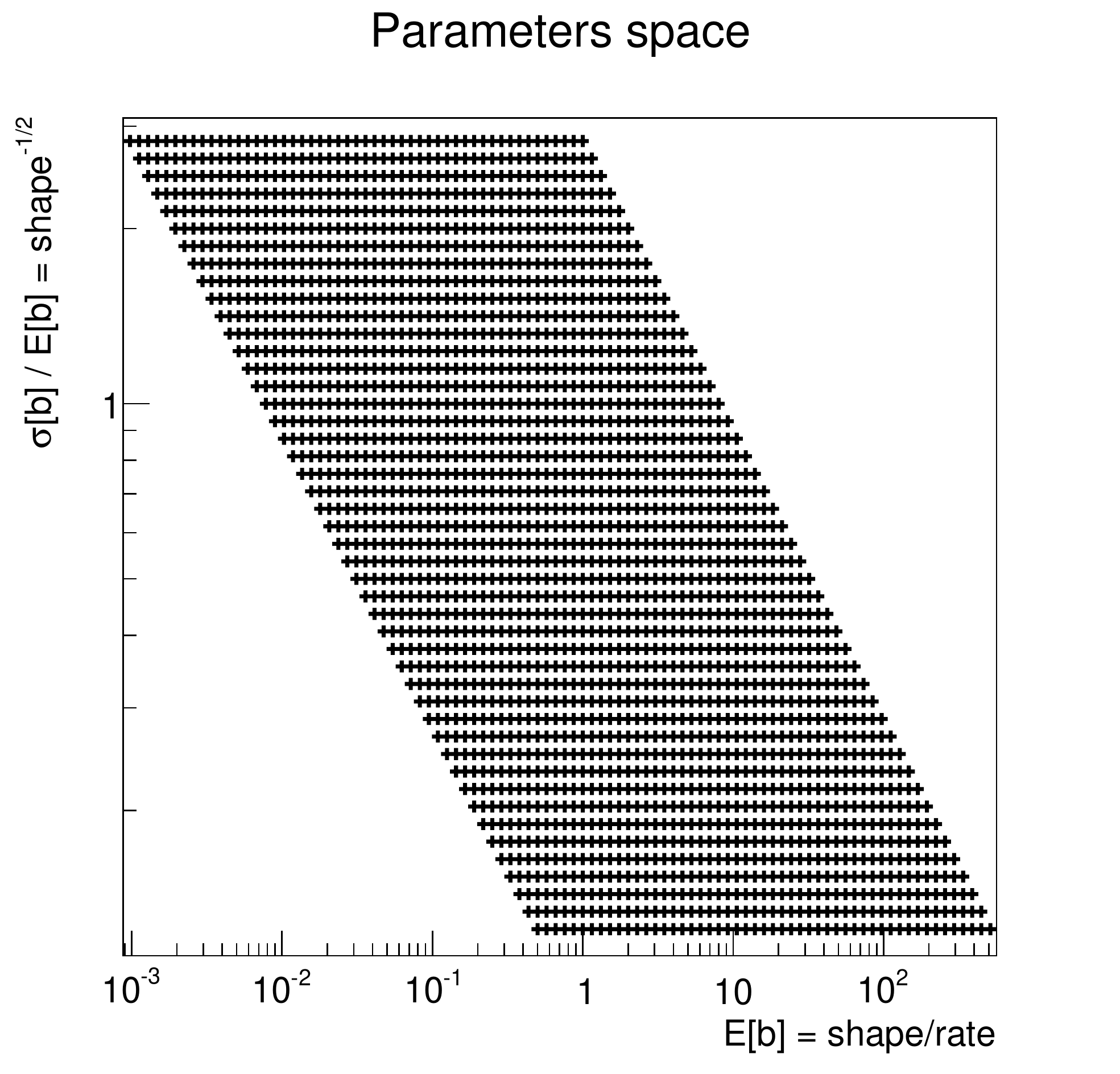}
   \end{minipage}%
   \caption{Scan of the background prior parameters space (left) and
     corresponding prior expectation and relative uncertainty (right).}
   \label{fig-param-space}
 \end{figure*}

 In order to see how well the limiting reference prior
 \(
    \pi_0(s) = \sqrt{E[b]/(s+E[b])}
 \)
 approximates the reference prior $\pi(s)$, the function $\pi_0(s)$
 has been computed over many points in the $(\alpha,\beta)$ parameter
 space.  A uniform logarithmic scan was performed,
 with a factor of 2 spanned by each parameter every 5 steps.
 Figure~\ref{fig-param-space} shows the points of the parameters
 space $(\alpha,\beta)$ which have been used in this paper, together
 with the corresponding values of the expected background
 $E[b]=\alpha/\beta$ and relative uncertainty
 $\sigma[b]/E[b]=1/\sqrt{\alpha}$.

\begin{figure*}[t!]
   \begin{minipage}{0.5\linewidth}
     \centering
     \includegraphics[width=0.95\textwidth]{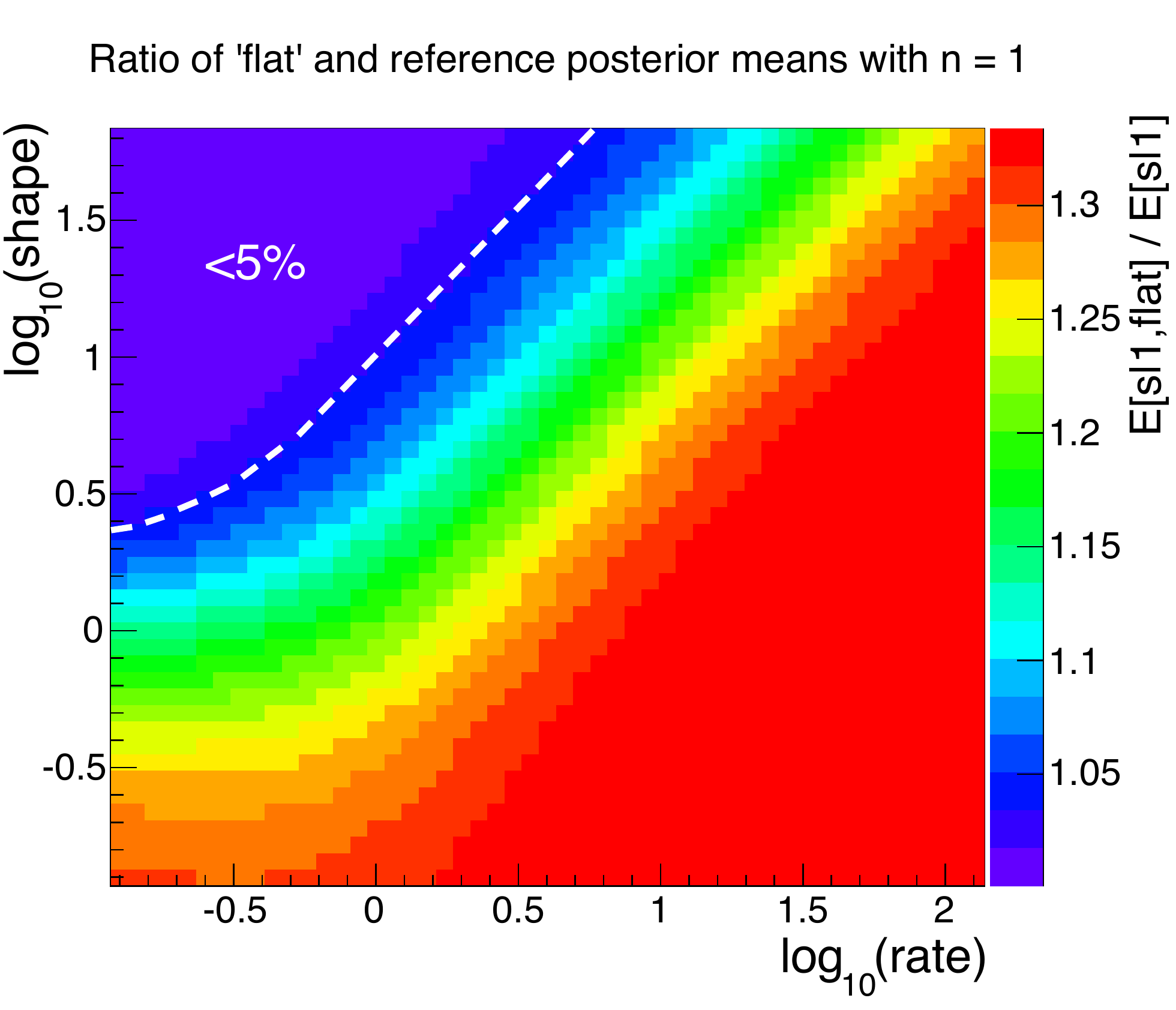}
   \end{minipage}%
   \begin{minipage}{0.5\linewidth}
     \centering
     \includegraphics[width=0.95\textwidth]{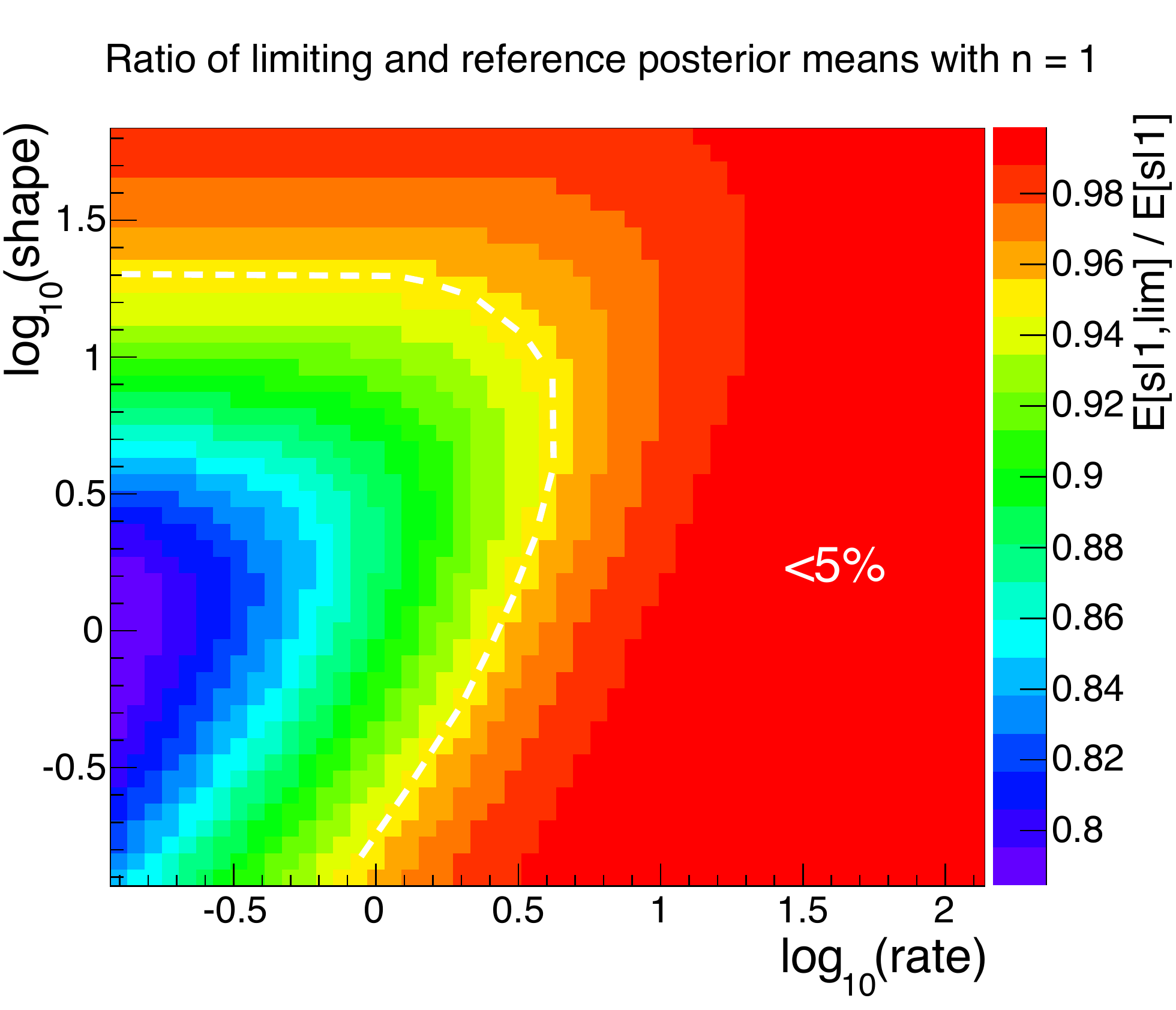}
   \end{minipage}%
   \caption{Ratio of posterior means for $n=1$ observed counts.  Left:
     the posterior corresponding to the flat prior $p(s)=1$ is
     compared to the reference posterior $p(s|n=1)$.  Right: the
     limiting reference posterior obtained when using the limiting
     prior $\pi_0(s)$ is compared to the reference posterior.}
   \label{fig-ratio-post-means}
 \end{figure*}

 The most useful comparison is among the posteriors obtained with
 different priors, and is definitely a very good check for any
 particular problem under analysis.  However, the posterior is
 conditional on the number $n$ of observed counts, which can range
 from zero to infinity, hence one would need to consider a very large
 number of possible combinations, which is unfeasible.
 Figure~\ref{fig-ratio-post-means} provides just one example, showing
 the ratio between the posterior expected signal computed with the
 flat or limiting prior and the reference posterior mean, when $n=1$
 (very far from the asymptotic regime, in which we know that the
 posteriors will be practically the same).  The flat prior always
 overestimates the posterior mean, approaching the reference posterior
 expectation when the background expectation is higher (i.e.~toward
 the top-left region in the parameters space).  This is the obvious
 consequence of assigning a costant weight to all possible values of
 the signal (in contrast, the reference prior assigns a monotonically
 decreasing weight to larger signal intensities).  On the other hand,
 the limiting reference prior expectation approaches the reference
 posterior mean from below, being closer than the result obtained with
 a flat prior over a large portion of the parameters space.  As an
 example, the regions in which the agreement is not worse than 5\% are
 also shown in the figure.

 As it is impractical to compare the posteriors for various values of
 $n$, we focus in the following on the differences between  priors.
 There are several ways of quantifying the ``distance'' between two
 probability distributions.  Among the most common ones, we have
 \begin{equation}
   \label{eq-Lp-distance}
   d_a = \left( \int |p(x)-q(x)|^a \di x \right)^{\!1/a}
 \end{equation}
 where $a=1$ corresponds to the area bracketed by the two densities
 $p(x)$ and $q(x)$, $a=2$ is the usual RMS difference, and
 $a\to\infty$ gives $d_\infty = \max |p(x)-q(x)|$, the maximum of the
 point-wise comparison.

\begin{figure*}[t!]
   \begin{minipage}{0.5\linewidth}
     \centering
     \includegraphics[width=\textwidth]{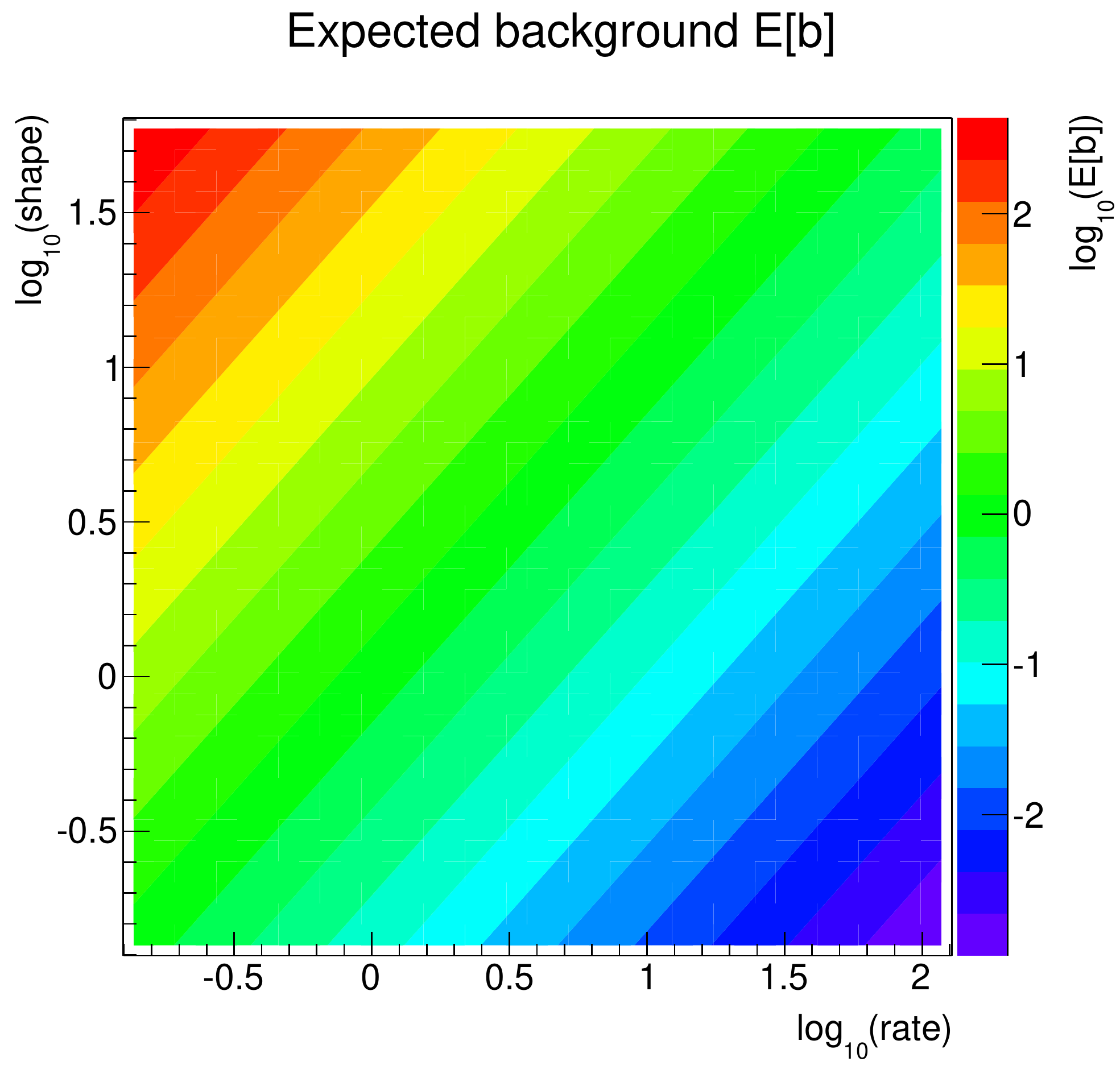}
   \end{minipage}%
   \begin{minipage}{0.5\linewidth}
     \centering
     \includegraphics[width=\textwidth]{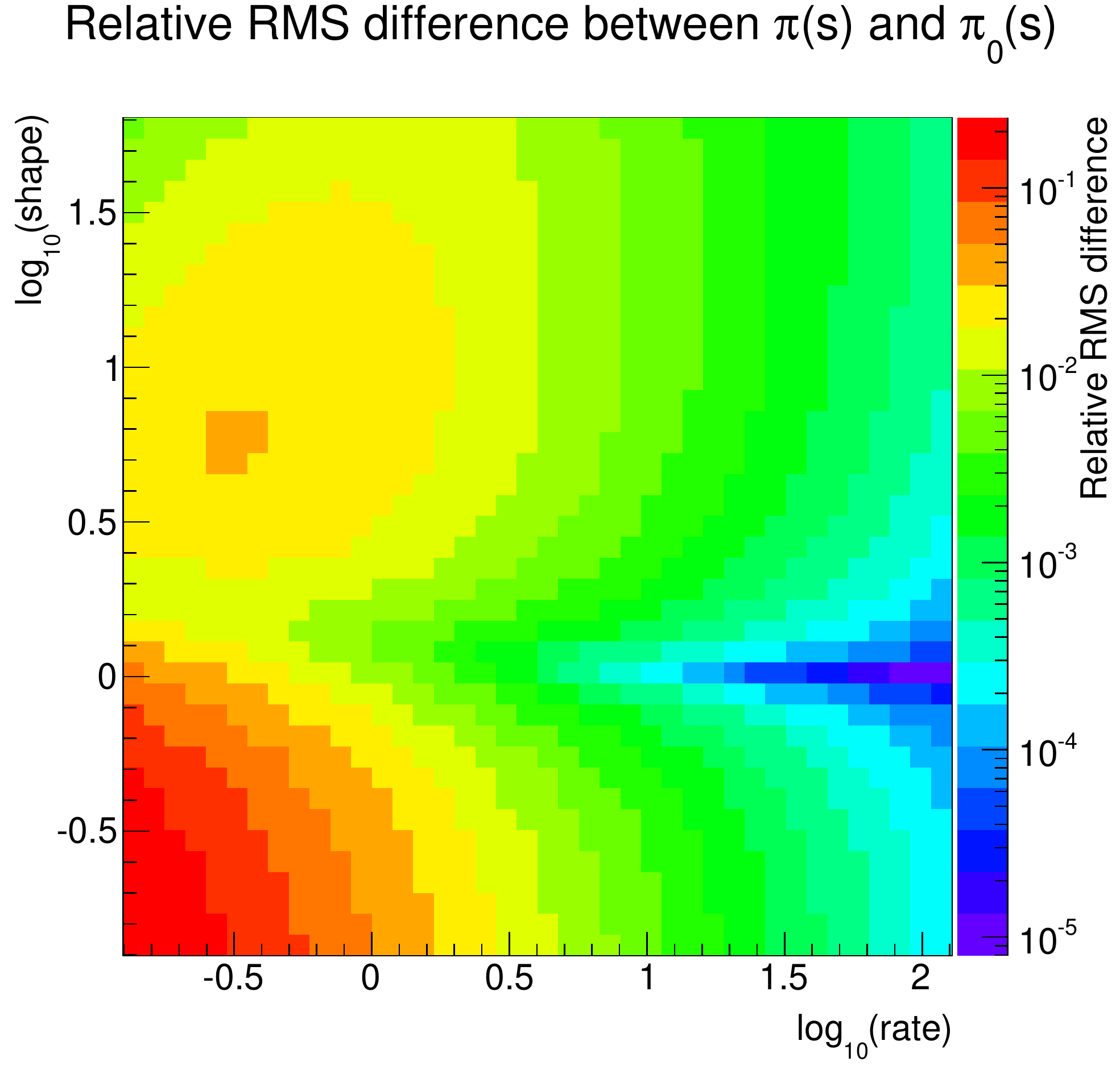}
   \end{minipage}%
   \caption{Expected background (left) and relative RMS difference
     between $\pi(s)$ and $\pi_0(s)$ (right).}
   \label{fig-f0}
 \end{figure*}

 All of them are defined for proper densities, whereas in our case we
 have improper priors to be compared.  Here we avoid problems related
 to the integration over an unlimited domain by restricting it to a
 reasonably wide range, chosen to be $[0,70]$.\footnote{The right edge
   is a compromise between computation effectiveness and rounding
   errors in the evaluation of $\pi(s)$.}  In addition, we choose
 the RMS distance $d_2$ and normalize the difference by dividing by the
 integral of the reference prior over the same interval.
 Figure~\ref{fig-f0} shows the expected background and the relative
 RMS difference between $\pi_0(s)$ and $\pi(s)$, as a function of
 $\alpha$ and $\beta$.

 In practice, the use of the reference prior is recommended mostly
 when the signal is small or absent: what really matters is the
 difference between the posteriors, and this becames negligible very
 quickly for increasing number $n$ of observed counts.  The dependence
 on the choice of the prior is strongest when $n$ is very small or
 zero, which implies that the sum of signal and background is very
 small.  In addition, here we are comparing monotonically decreasing
 functions all starting from the same positive value (one at the
 origin) and all asymptotically reaching zero: their relative
 difference (apart from the maximum point-wise separation) becomes
 smaller and smaller for increasing values of the argument.  This
 means that, although somewhat arbitrary, the choice of looking at the
 relative RMS difference over $[0,70]$ is sufficient to capture all
 relevant aspects of the comparison and can serve as a useful
 guideline for deciding when one should use the full reference prior
 or can safely use the approximate expression.


 For most practical purposes, a relative RMS difference below 1\% is
 acceptable, as this is the order of magnitude of the maximum change
 in the posterior in the limit of very few or zero observed counts.
 Hence we will assume that a disagreement not larger than 1\% is
 tolerable in this paper.  On the other hand, it should be emphasized
 that a threshold at 1\% is quite conservative.  In most applications
 larger deviations can be tolerated, as the posteriors will quickly
 become indistinguishable for increasing number $n$ of observed
 counts.  In addition, the common practice is to summarize the
 posterior by providing one value (e.g.~the expectation)
 and some estimate of its uncertainty (e.g.~the shortest interval
 covering 68.3\% posterior probability), by rounding the values to the
 minimum meaningful number of digits.  Often, this summary is quite
 robust compared to relative RMS differences of several percent.
 Nevertheless, the notion of ``acceptable difference'' is application
 dependent, and the researcher should state which is the criterion
 dictating the choice of the prior.

 The right panel of figure~\ref{fig-f0} shows that the
 limiting reference prior is satisfactory (differing by less than 1\%)
 when the rate parameter is larger than 4, and in some case even for
 lower values (depending on the shape parameter).


 \section{Computing the reference prior}

\begin{figure*}[t!]
   \begin{minipage}[b]{0.5\linewidth}
     \centering
   \includegraphics[width=\linewidth]{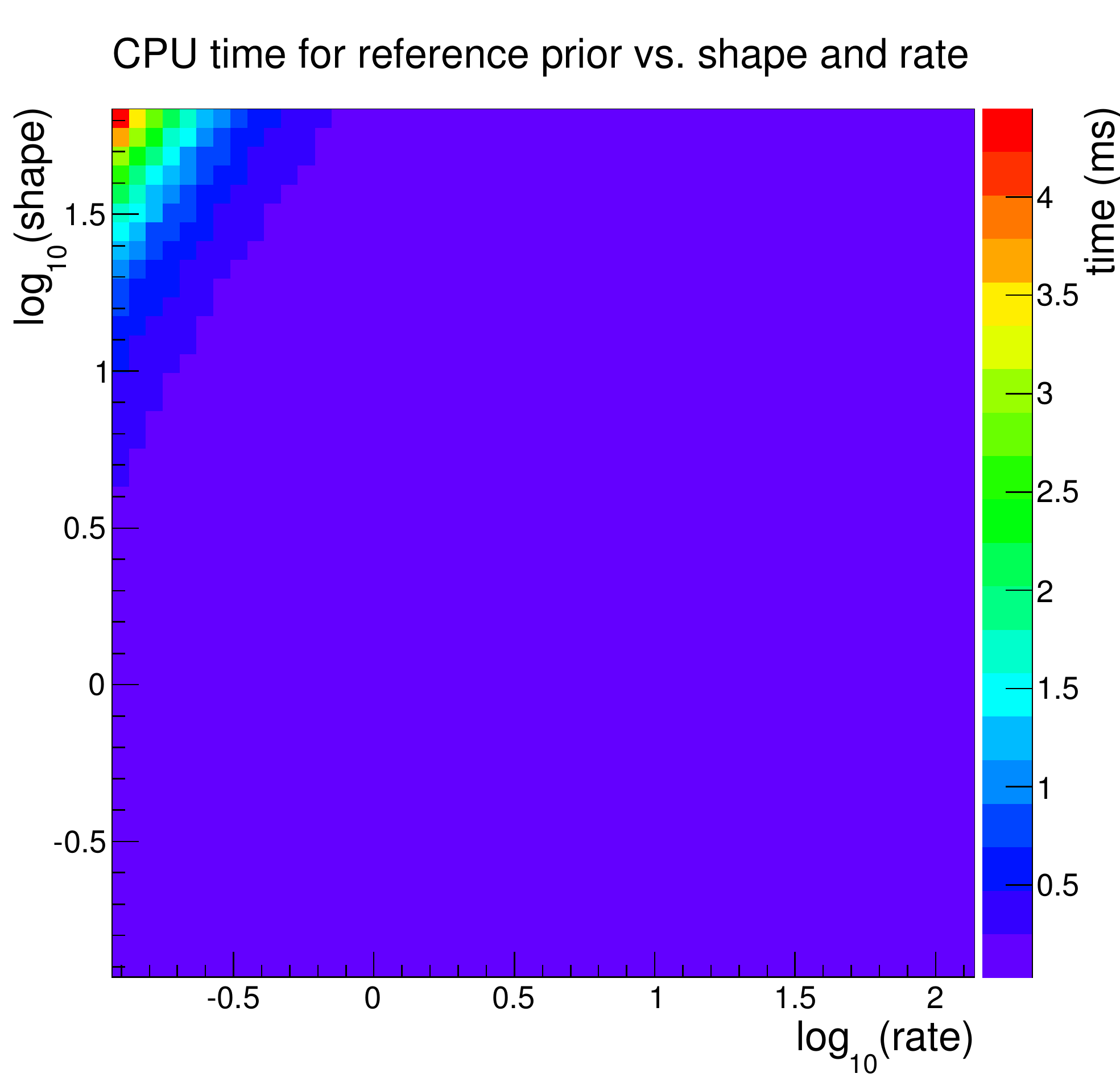}
   \end{minipage}%
   \begin{minipage}[b]{0.5\linewidth}
   \caption{Average CPU time per call required by the author's private
     C++ code computing the reference prior $\pi(s)$.} 
   \label{fig-cpu}
   \end{minipage}%
 \end{figure*}

 The reference prior (\ref{eq-ref-prior}) can be evaluated
 iteratively, by computing the next element of its infinite series
 until the difference with respect to the previous result becomes
 smaller than a predefined tolerance.  When $s$ is large, more terms
 need to be computed to achieve a predefined precision, which means
 that the calculation become slower.  In addition, the CPU time may
 depend on the background parameters.  For example the author's
 private C++ code works with a tolerance of $10^{-6}$, a bit larger
 than the accuracy of single-precision floating-point values.  The
 average CPU time per evaluated point with $s\in[0,70]$ ranges from
 0.5~ms to 4.5~ms on the author's laptop, with the lowest value
 spanning over most of the parameters space, apart from the top-left
 corner (shape higher than 10, rate lower than 1;
 figure~\ref{fig-cpu}), where it increases rapidly.

 With the goal of improving the performance of numerical evaluations,
 approximations of the reference prior were searched for, which could
 provide better results than the limiting reference prior
 (\ref{eq-lim-ref-prior}).  Closed form expressions were found, which
 allow to significantly reduce the computing time.  The 2-parameter
 approximation illustrated in section~\ref{sec-f2} below requires
 0.10--0.24~\micro{s} per call, independently of the region in the
 $(\alpha,\beta)$ space, providing a speed-up in the range
 $10^3$--$2\times10^4$.

 \subsection{A 1-parameter approximation}\label{sec-f1}

\begin{figure*}[t!]
   \begin{minipage}{0.5\linewidth}
     \centering
     \includegraphics[width=\textwidth]{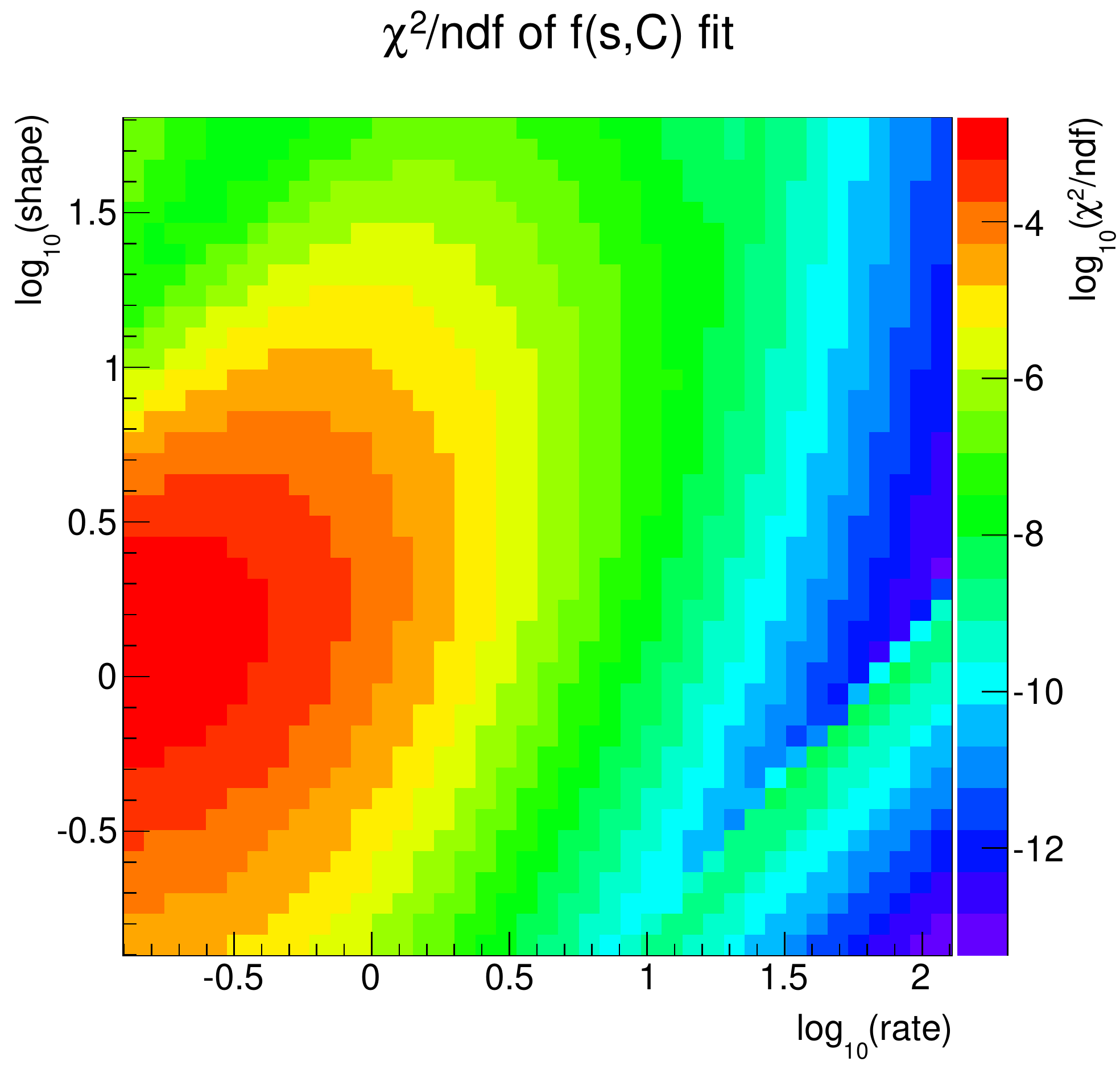}
   \end{minipage}%
   \begin{minipage}{0.5\linewidth}
     \centering
     \includegraphics[width=\textwidth]{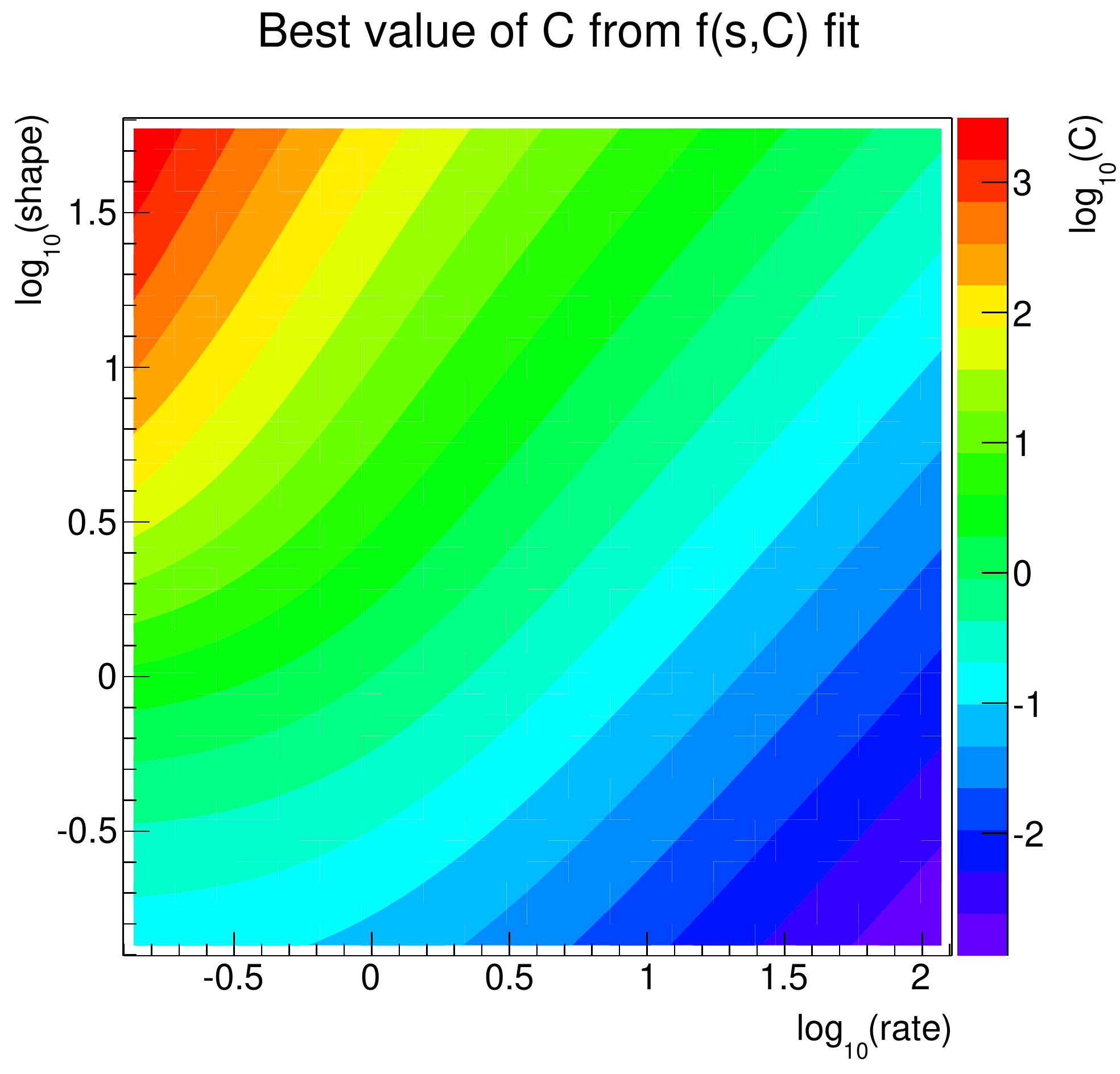}
   \end{minipage}%
   \caption{Fit quality (left) and best parameter value (right).}
   \label{fig-chi2ndf-C}
 \end{figure*}

\begin{figure*}[t!]
   \begin{minipage}{0.5\linewidth}
     \centering
     \includegraphics[width=\textwidth]{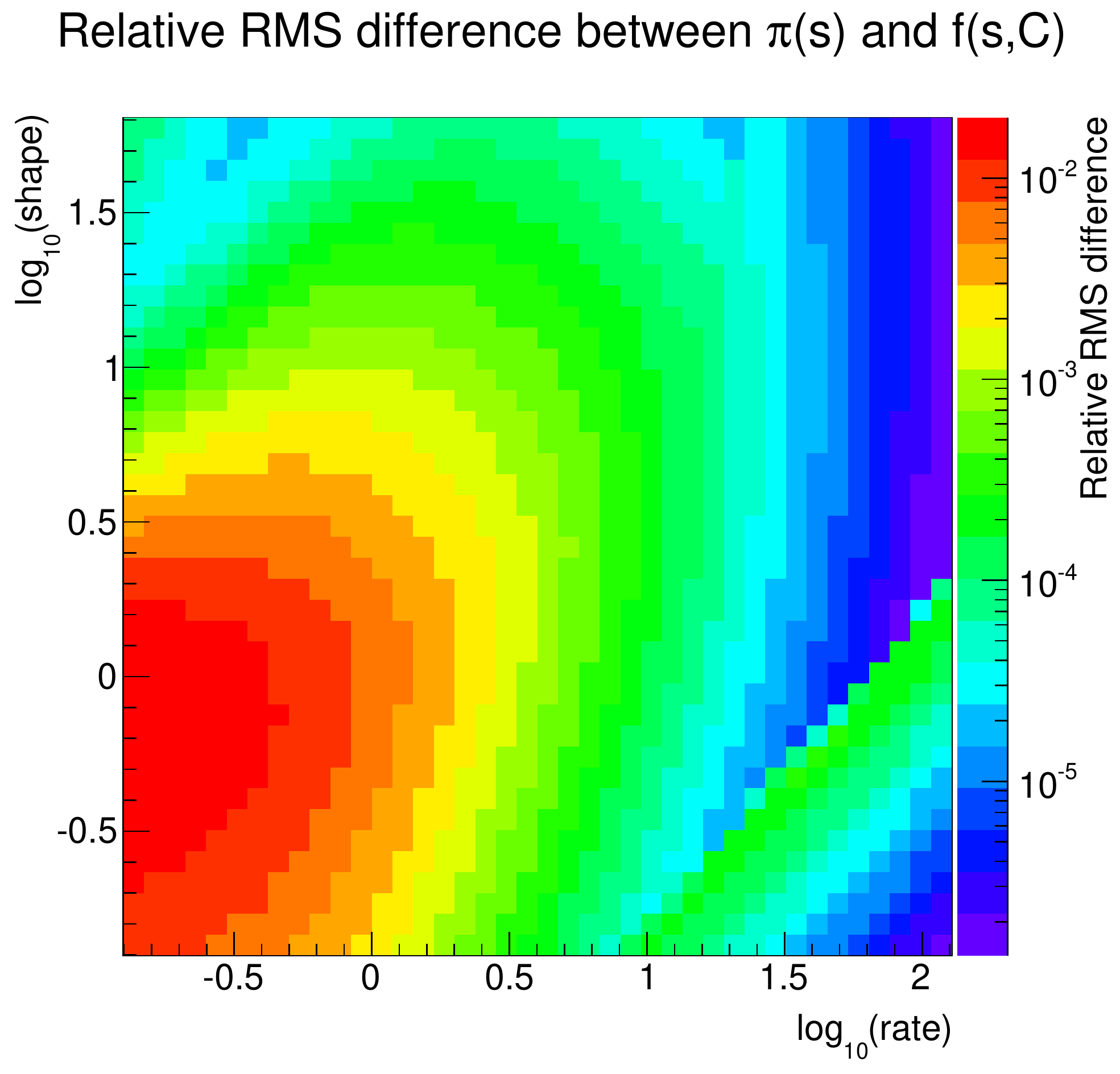}
   \end{minipage}%
   \begin{minipage}{0.5\linewidth}
     \centering
     \includegraphics[width=\textwidth]{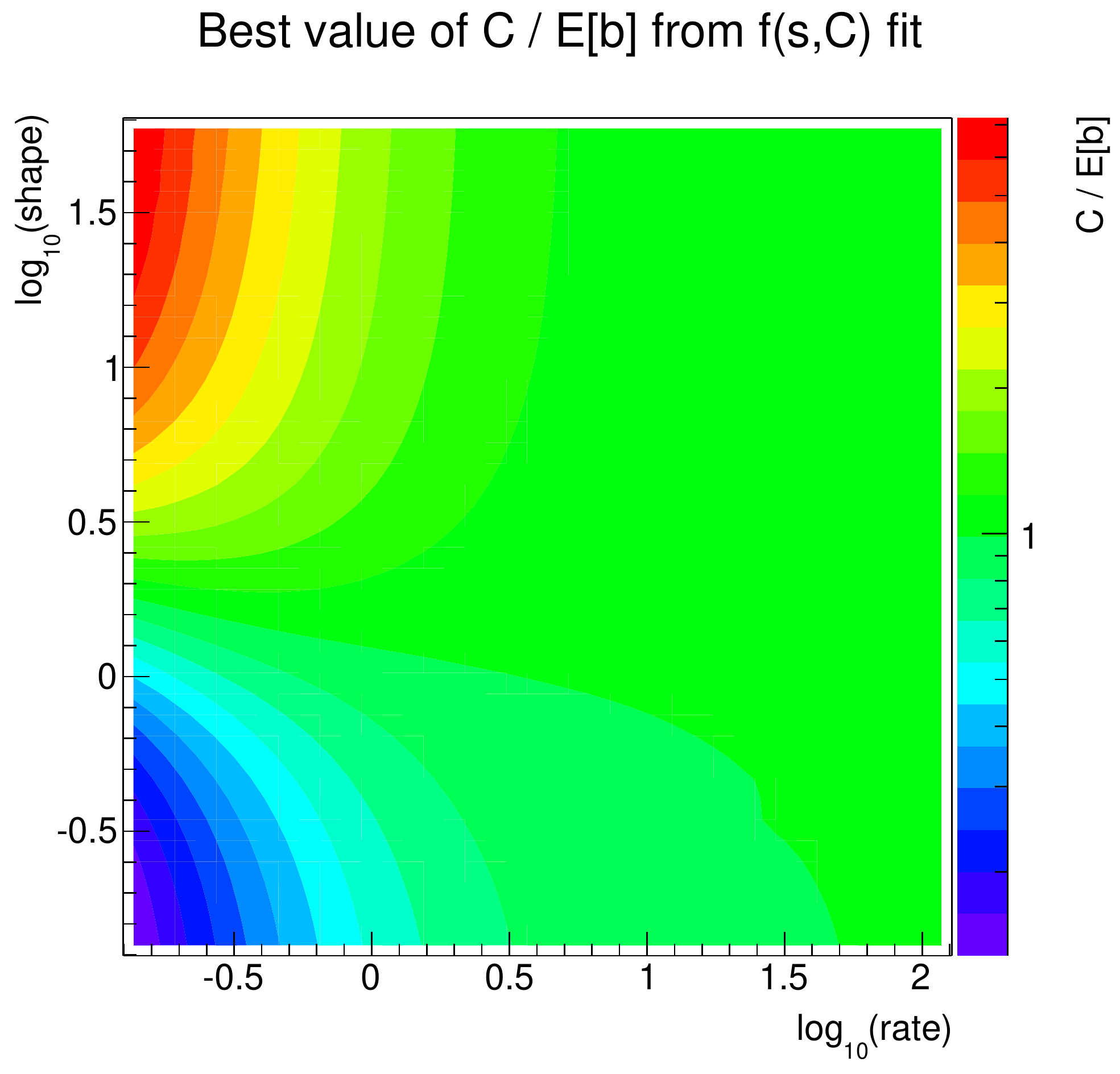}
   \end{minipage}%
   \caption{Relative RMS difference between the best fitting function
     $f(s; C)$ and the reference prior $\pi(s)$ (left), and ratio
     between the fit parameter and the prior background expectation
     (right).}
   \label{fig-RMS-and-CdivE}
 \end{figure*}

 Inspired by the limiting form (\ref{eq-lim-ref-prior}), one may look
 for a simple approximation of $\pi(s)$ in which a single parameter is
 tuned to obtain the best agreement with the reference prior.  A fit
 has been performed with the function
 \begin{equation}\label{eq-fit-func1}
    f(s; C) = \sqrt{\frac{C}{s+C}}
 \end{equation}
 where $C>0$ is the single unknown parameter, over the parameters
 space.  The value of the reference prior $\pi(s)$ has been computed
 at equally spaced steps\footnote{The data file is freely available;
   see Appendix~\ref{app-data}.} in the range $s\in[0,70]$.  The fit
 quality and best parameter values are shown in
 figure~\ref{fig-chi2ndf-C}.  Clearly, $C\to E[b]$ in the limit of
 perfect background knowledge, where $\pi_0(s)$ works well
 (figure~\ref{fig-RMS-and-CdivE}, right panel).

 The relative RMS difference between $f(s;C)$ and $\pi(s)$ is shown in
 figure~\ref{fig-RMS-and-CdivE}, left panel.  The function
 (\ref{eq-fit-func1}) provides a good approximation, with a relative
 RMS difference below 1\%, whenever both shape and rate parameters are
 not small (i.e.~when they are at least few units).  Our quality
 threshold is exceeded only if $\alpha<2.5$ \emph{and} $\beta<0.6$.

 When there is very good prior knowledge of the background, the
 limiting value for the $C$ parameter is the prior background
 expectation, hence it is instructive to look at the ratio $C/E[b]$
 over the parameter space.  As the limiting situation which we are
 considering involves a fixed background expectation and a decreasing
 relative precision $\sqrt{V[b]}/E[b] = 1/\sqrt{\alpha}$, we expect
 that far from the conditions of perfect background knowledge the
 departure of $C/E[b]$ will depend on the shape parameter only.
 Indeed, the right plot in figure~\ref{fig-RMS-and-CdivE} shows that
 for small rate values, the ratio is a monotonically increasing
 function of the shape parameter, whereas for large values of the rate
 parameter (in practice, when $\beta$ is at least about ten), $C$ is
 practically equivalent to the prior background
 expectation. 
   In particular, the logarithm of the ratio
   $C/E[b]$ can be well fitted by an arcotangent function of the
   logarithm of the shape parameter, whose amplitude (the distance
   between the two asymptotic values) goes to zero very quickly with
   increasing $\beta$ values (a Gaussian fit well reproduces the
   amplitude as a function of the logarithm of the rate parameter in
   the range studied here).

 In summary, when the rate parameter is larger than several units the
 limiting reference prior (\ref{eq-lim-ref-prior}) already provides a
 good approximation.  In addition, for smaller values of $\beta$ the
 1-parameter function (\ref{eq-fit-func1}) can fit the reference prior
 well enough, provided that the shape parameter is larger than few
 units.  However, this approximation does not satisfy our requirement
 over the entire parameter space, hence we look for something better.


 \subsection{A 2-parameter approximation}\label{sec-f2}

\begin{figure*}[t!]
   \begin{minipage}{0.5\linewidth}
     \centering
     \includegraphics[width=\textwidth]{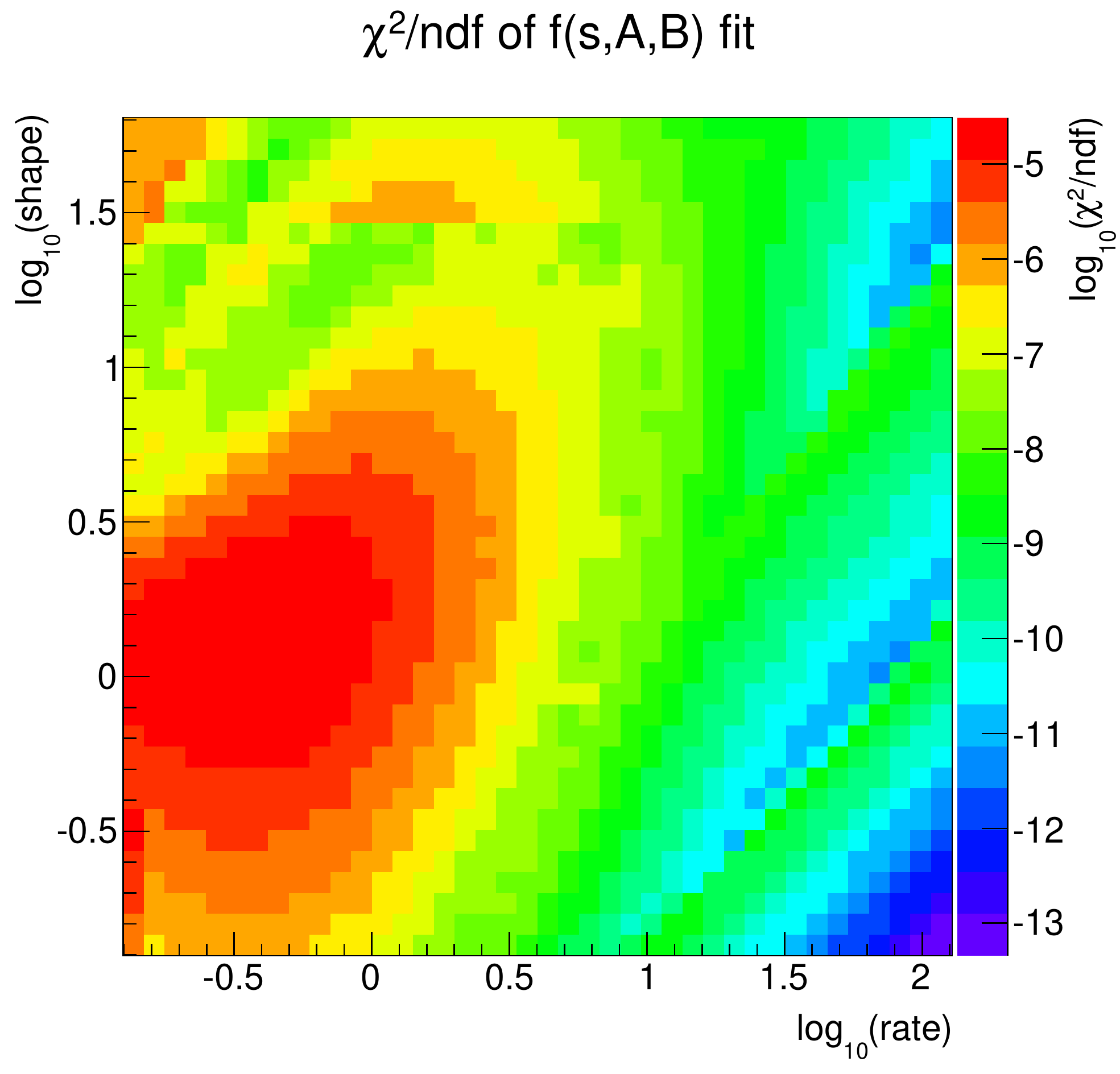}
   \end{minipage}%
   \begin{minipage}{0.5\linewidth}
     \centering
     \includegraphics[width=\textwidth]{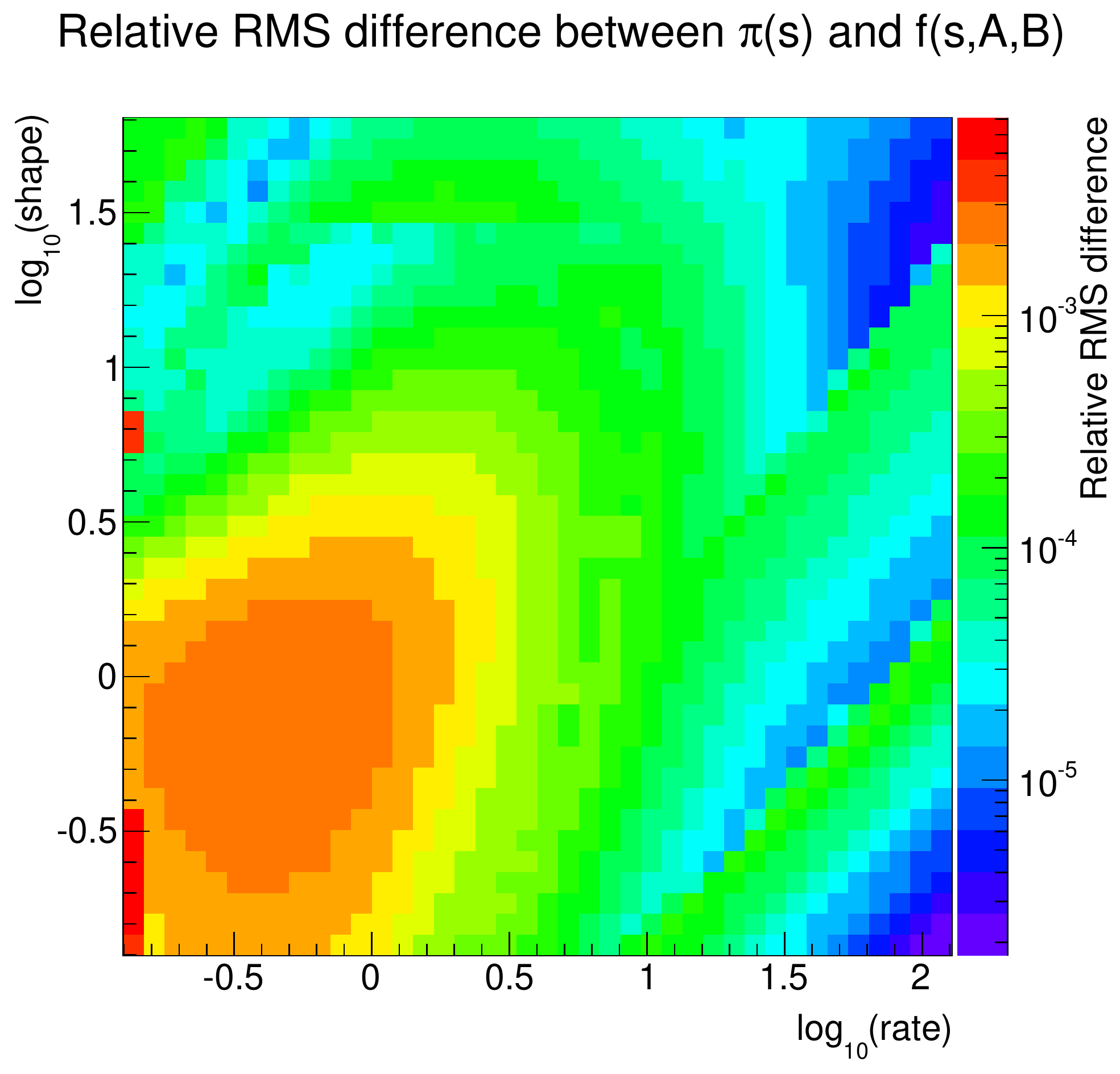}
   \end{minipage}%
   \caption{Fit quality (left) and relative RMS difference between the
     best fitting function $f(s; A,B)$ and the reference prior
     $\pi(s)$ (right).}
   \label{fig-AB-chi2-RMS}
 \end{figure*}

\begin{figure*}[t!]
   \begin{minipage}{0.5\linewidth}
     \centering
     \includegraphics[width=\textwidth]{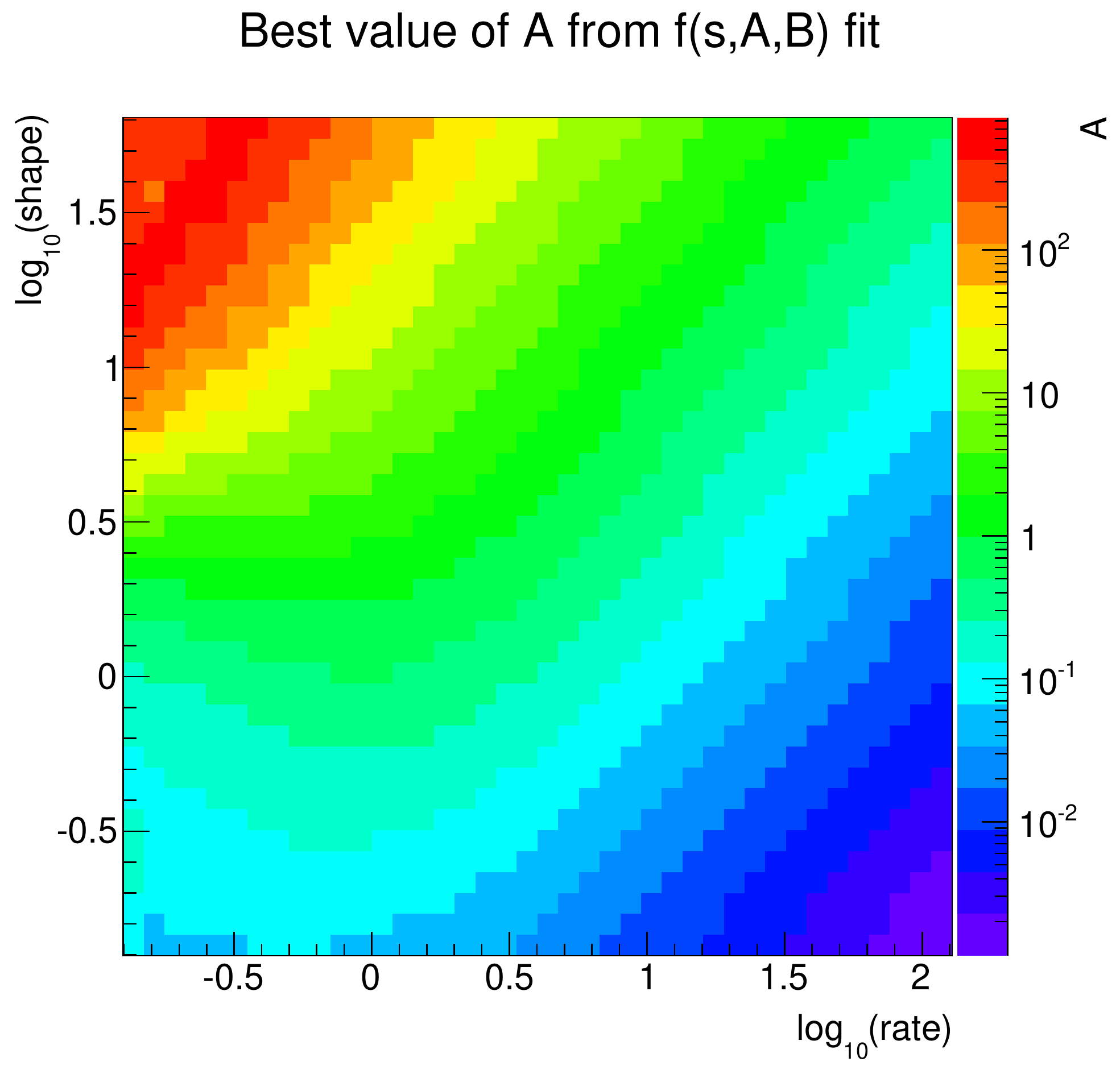}
   \end{minipage}%
   \begin{minipage}{0.5\linewidth}
     \centering
     \includegraphics[width=\textwidth]{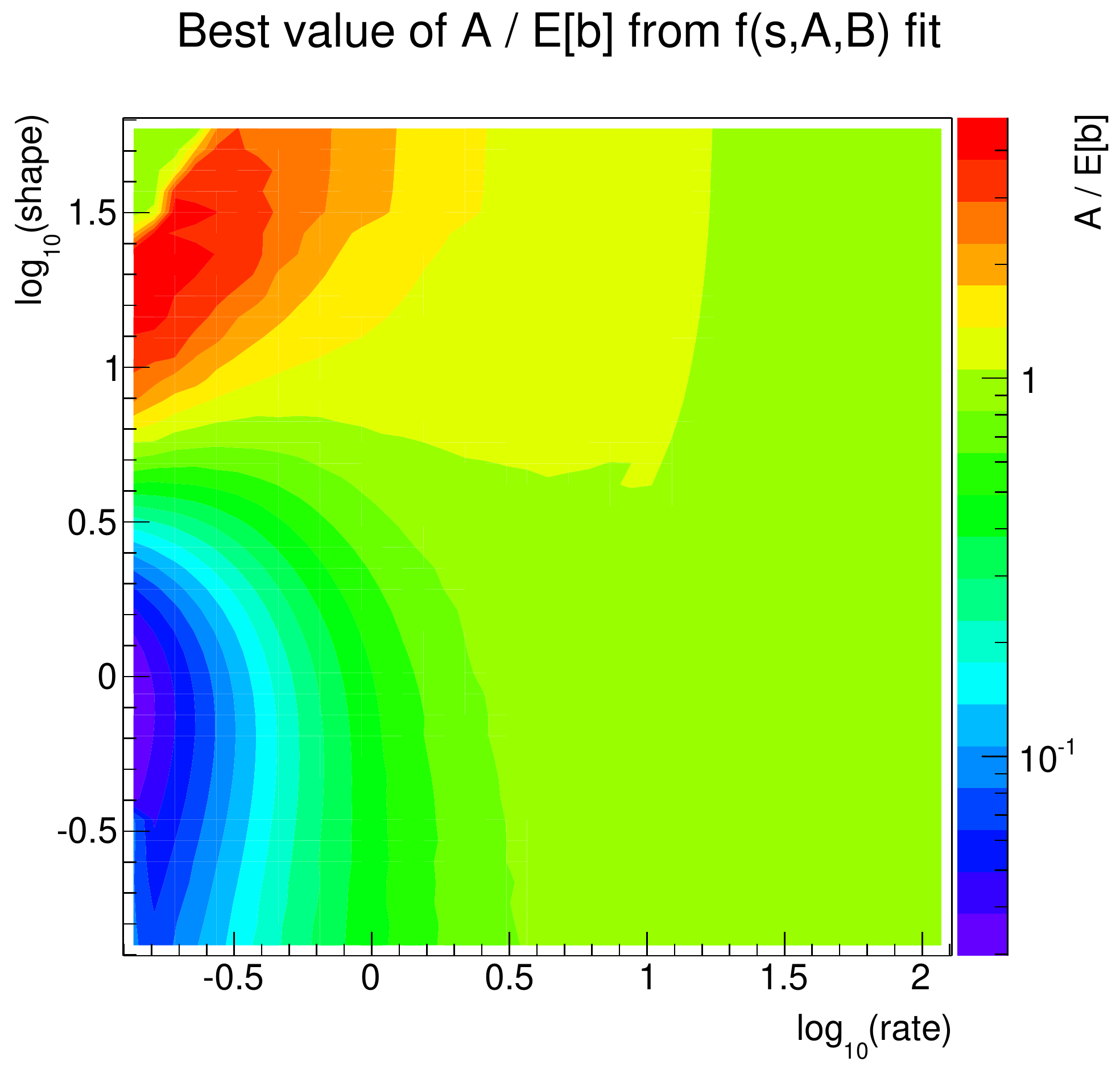}
   \end{minipage}%
   \caption{Value of the $A$ parameter for the best fitting function
     $f(s; A,B)$ (left), and its ratio with the prior background
     expectation (right).}
   \label{fig-AB-A}
 \end{figure*}

\begin{figure*}[t!]
   \begin{minipage}[b]{0.5\linewidth}
     \centering
     \includegraphics[width=\textwidth]{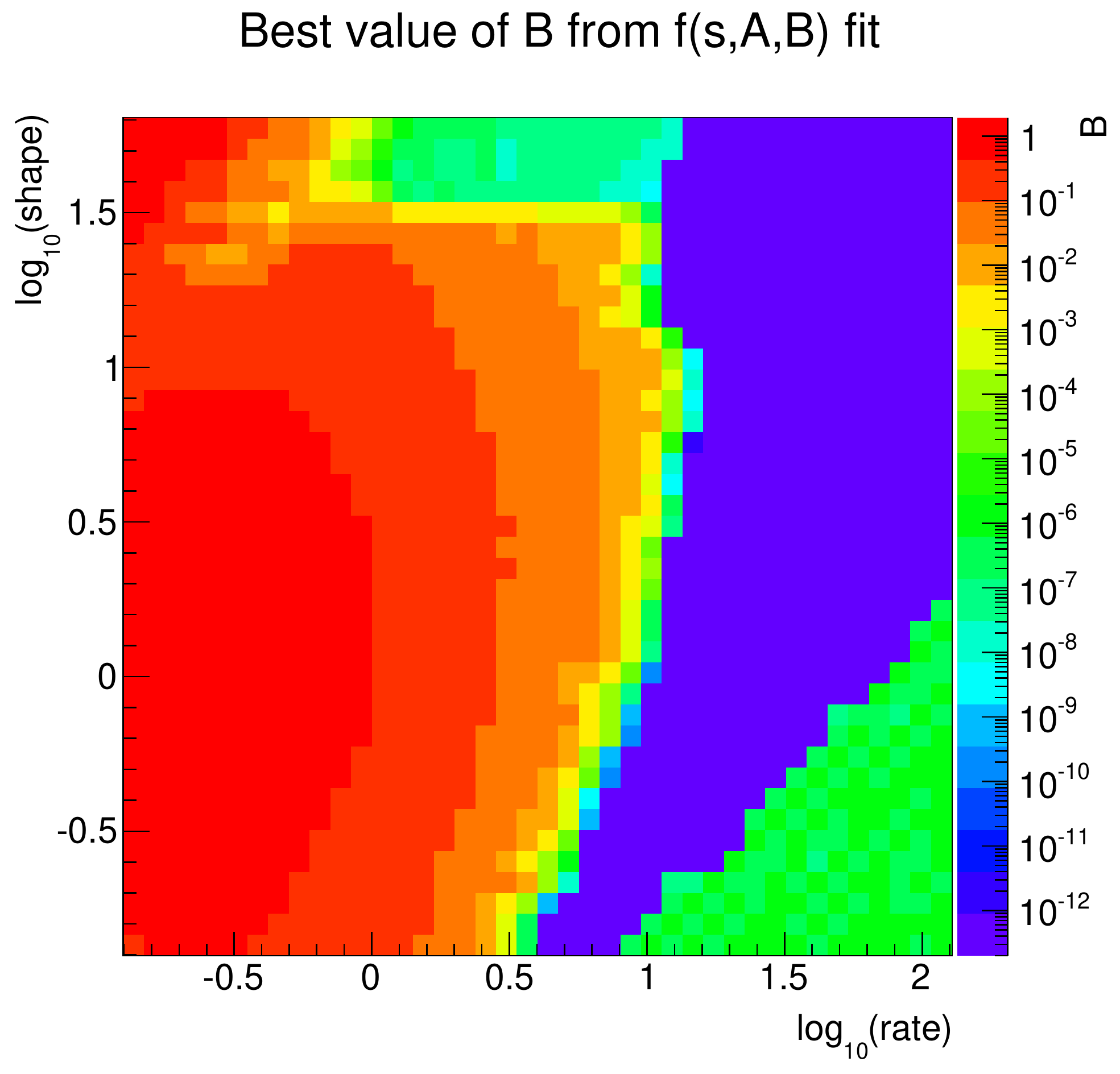}
   \end{minipage}%
   \begin{minipage}[b]{0.5\linewidth}
   \caption{Value of the $B$ parameter for the best fitting function
     $f(s; A,B)$.}
   \label{fig-AB-B}
   \end{minipage}%
 \end{figure*}

 A function which fits the reference prior better than
 (\ref{eq-fit-func1}) and can well reproduce $\pi(s)$ over
 the entire portion of parameters space considered here is
 \begin{equation}\label{eq-fit-func2}
    f(s; A,B) = \sqrt{\frac{A \exp(B\,s^{E})}{s+A \exp(B\,s^{E})}}
 \end{equation}
 It coincides with (\ref{eq-fit-func1}) when setting the parameter $D$
 to zero and should be used if $\alpha<2.5$ \emph{and} $\beta<0.6$
 \emph{and} $n$ is small.
 The power $E$ of $s$ in the exponent was found not to change
 appreciably over the entire parameter space, and was treated as a
 fixed parameter with value $E=0.125$ when producing the plots shown
 in figures \ref{fig-AB-chi2-RMS}, \ref{fig-AB-A} and \ref{fig-AB-B}
 (only $A$ and $B$ were left free to vary in the fits).
 The difference with the reference prior is very small and completely
 negligible in any practical application.\footnote{Very similar
   results are obtained for few percent changes in the $E$ parameter,
   with differences which might be influenced by rounding errors.  The
   entire suite of fits was performed with $E=0.100$, obtaining
   essentially the same fit quality everywhere.}

 Figure~\ref{fig-AB-chi2-RMS} shows the goodness of fit of
 (\ref{eq-fit-func2}) and the relative RMS difference between $f(s;
 A,B)$ and $\pi(s)$.  The latter is one order of magnitude better than
 the result obtained with $f(s;C)$, and is always smaller that 1\% over
 the entire portion of the parameters space investigated here.  Hence
 the form (\ref{eq-fit-func2}) is well suited for practically all
 applications.
 Indeed, this function can be optionally used in BAT \cite{BAT2009} to
 speed-up the computation of the reference prior: the latter is
 initially computed over a number of discrete $s$ values, then a best
 fit with $f(s; A,B)$ is performed, and the latter is used in all
 following computations (a similar approach is illustrated in
 Appendix~\ref{app-data}).

 As shown in figure~\ref{fig-AB-A}, the best value of the $A$
 parameter basically coincides with the best value of $C$ whenever
 $f(s;C)$ represents a good approximation.  In this portion of the
 parameters space (say for $\beta$ above several units), the $B$
 parameter is indeed so small that one can round it off to zero
 (figure~\ref{fig-AB-B}).

 \section{Summary and conclusions}

 The reference prior $\pi(s)$ for the $\Poi(n|s+b)$ model can be
 computed when an informative prior for the nuisance parameter $b$ is
 available in the form of a Gamma density with known shape and rate
 parameters.  The reference prior is an improper density, as it can be
 expected by analogy with Jeffreys' prior for a single Poisson
 variable.  For practical applications, it is recommended to fix the
 arbitrary multiplicative constant in such a way that $\pi(s)$ is a
 monotonically decreasing function with maximum $\pi(0)=1$, as this
 simplifies the comparison with the widespread uniform prior.

 The limiting form of $\pi(s)$ when there is certain information about
 the backgrond is $\pi_0(s) = \sqrt{b_{0}/(s+b_{0})}$, which is
 Jeffreys' prior for the offset-ed variable $s' = s+b_{0}$.  The
 corresponding posterior $p_0(s|n) \propto \Ga(s+b_0|n+\frac{1}{2},1)$
 provides a valid approximation to the full reference posterior in
 many cases.  In particular, this is true when the relative
 uncertainty $\sigma[b]/E[b]=1/\sqrt{\alpha}$ on the background in the
 ``signal region'' is small, i.e.~for large values of the shape
 parameter $\alpha$.  In addition, even when $\alpha$ is small, the
 approximate prior $\pi_0(s)$ differs less than 1\% from $\pi(s)$ when
 the rate parameter $\beta$ is larger than few units.

 In most cases, $\pi_0(s)$ approximates $\pi(s)$ much better than the
 uniform prior and the resulting posterior is much simpler.  As the
 Gamma density is available in all software packages used in data
 analysis, evaluating the approximate reference posterior $p_0(s|n)$
 is straightforward.  When $n$ is not too small (in practice, a few
 counts are often sufficient), it will provide a very good
 approximation to the full reference posterior.  For these reasons, it
 is recommended to consider $\pi_0(s)$ as the best default or
 conventional prior, in place of the flat prior, whenever the use of
 the full reference prior $\pi(s)$ is considered too complicated.

 The user shall decide when an approximate form gives acceptable
 results in her application.  In this paper, we adopted an overall
 agreement not worse than 1\% between the priors as a (very
 conservative) guideline.  This implies that, when $\alpha<30$
 \emph{and} $\beta<4$, the full reference prior (or its 2-parameter
 approximation) should be used.  On the other hand, this rule ignores
 the fact that, unless $n$ is very low or zero, in practice the
 difference between the corresponding posteriors (the things which
 matter) is smaller than the difference between priors (and approaches
 zero when $n$ increases).

 The first public implementation of the reference posterior in BAT has
 also the option of reducing the computing power by means of the
 2-parameter function~(\ref{eq-fit-func2}), which can reproduce
 $\pi(s)$ over the entire range of parameters scanned in this work.
 This option might be useful in applications requiring the evaluation
 of the reference prior a very large number of times, as the speed-up
 is at least of three orders of magnitude.  People who do not use BAT
 may implement the same excellent approximation to the reference
 prior, which is used in BAT to speed-up the computation, by looking
 at the values of $\pi(s)$ computed for all the points in the
 background parameters space studied here, as explained in
 Appendix~\ref{app-data}.


 \section*{Acknowledgments}

 The author wishes to thank Kevin Kr\"{o}ninger for the numerous cross
 checks he performed while implementing in BAT \cite{BAT2009} the
 reference prior, including the approximation described here, and for
 providing the example reported in Appendix~\ref{sec-bat}.


 \appendix

 \section{Example with BAT}\label{sec-bat}

\begin{figure}[t]
   \begin{minipage}{0.5\textwidth}
     \centering
     \includegraphics[width=0.95\textwidth]{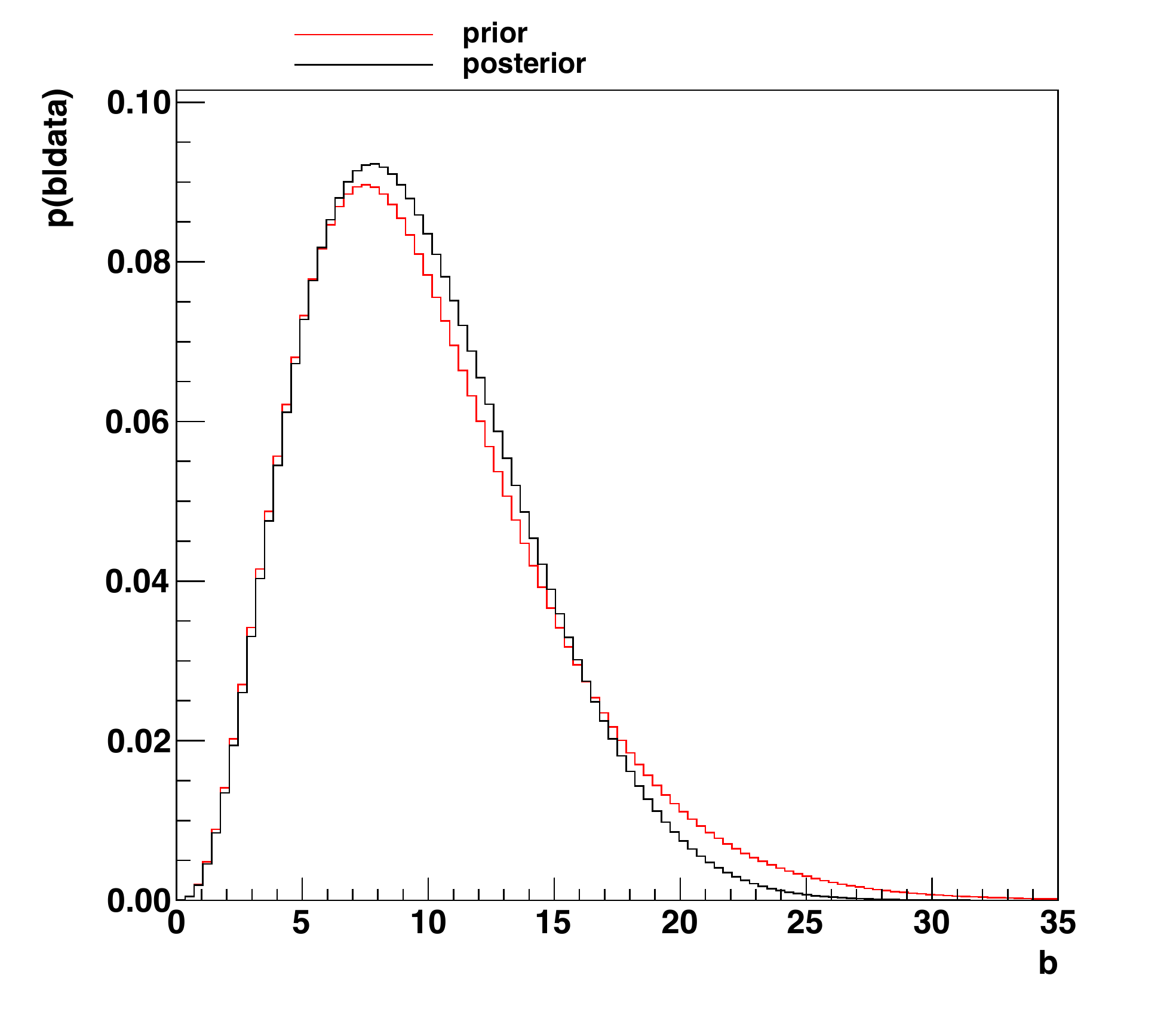}
   \end{minipage}%
   \begin{minipage}{0.5\textwidth}
     \centering
     \includegraphics[width=0.95\textwidth]{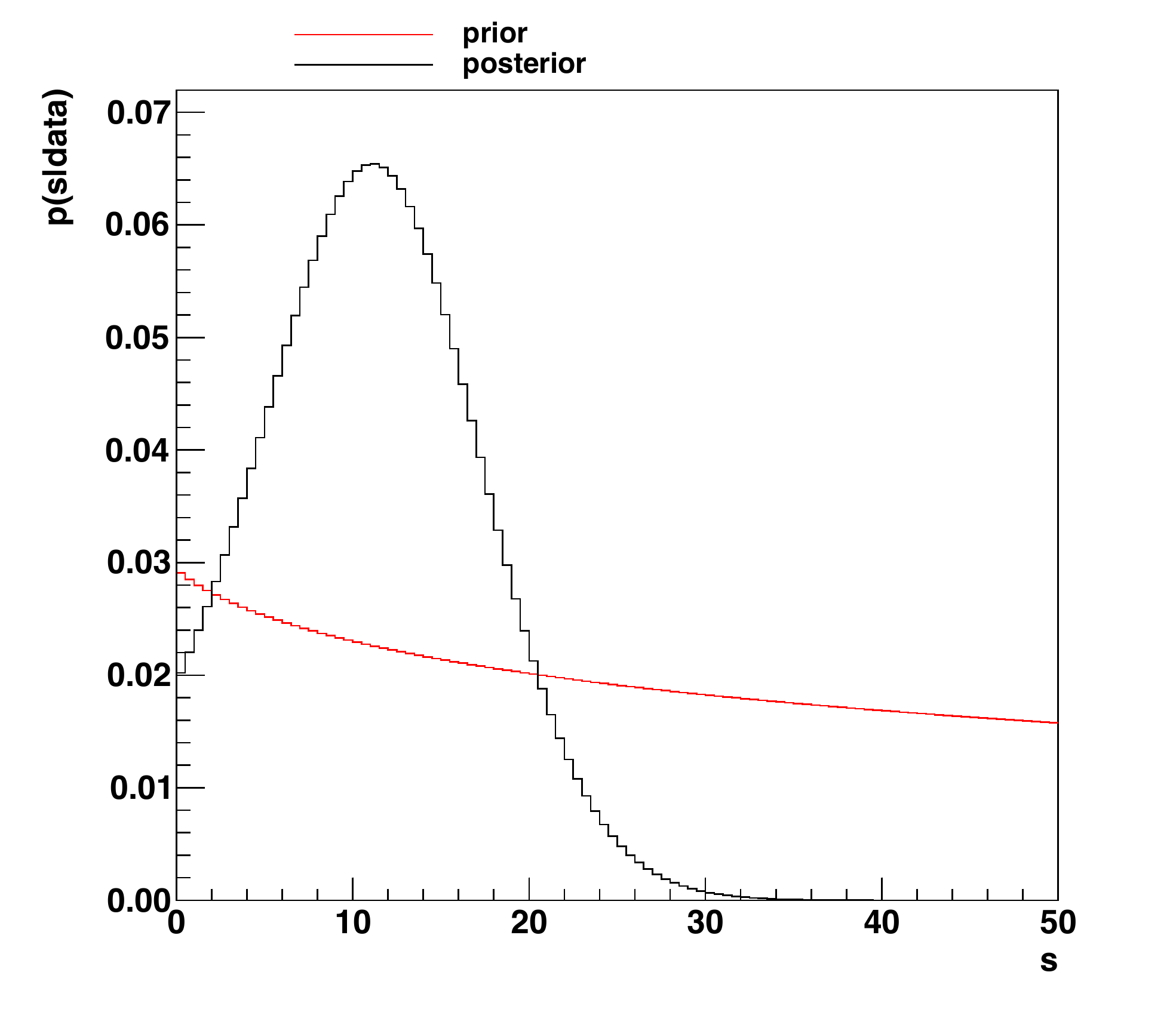}
   \end{minipage}

   \begin{minipage}{0.5\textwidth}
     \centering
     \includegraphics[width=0.95\textwidth]{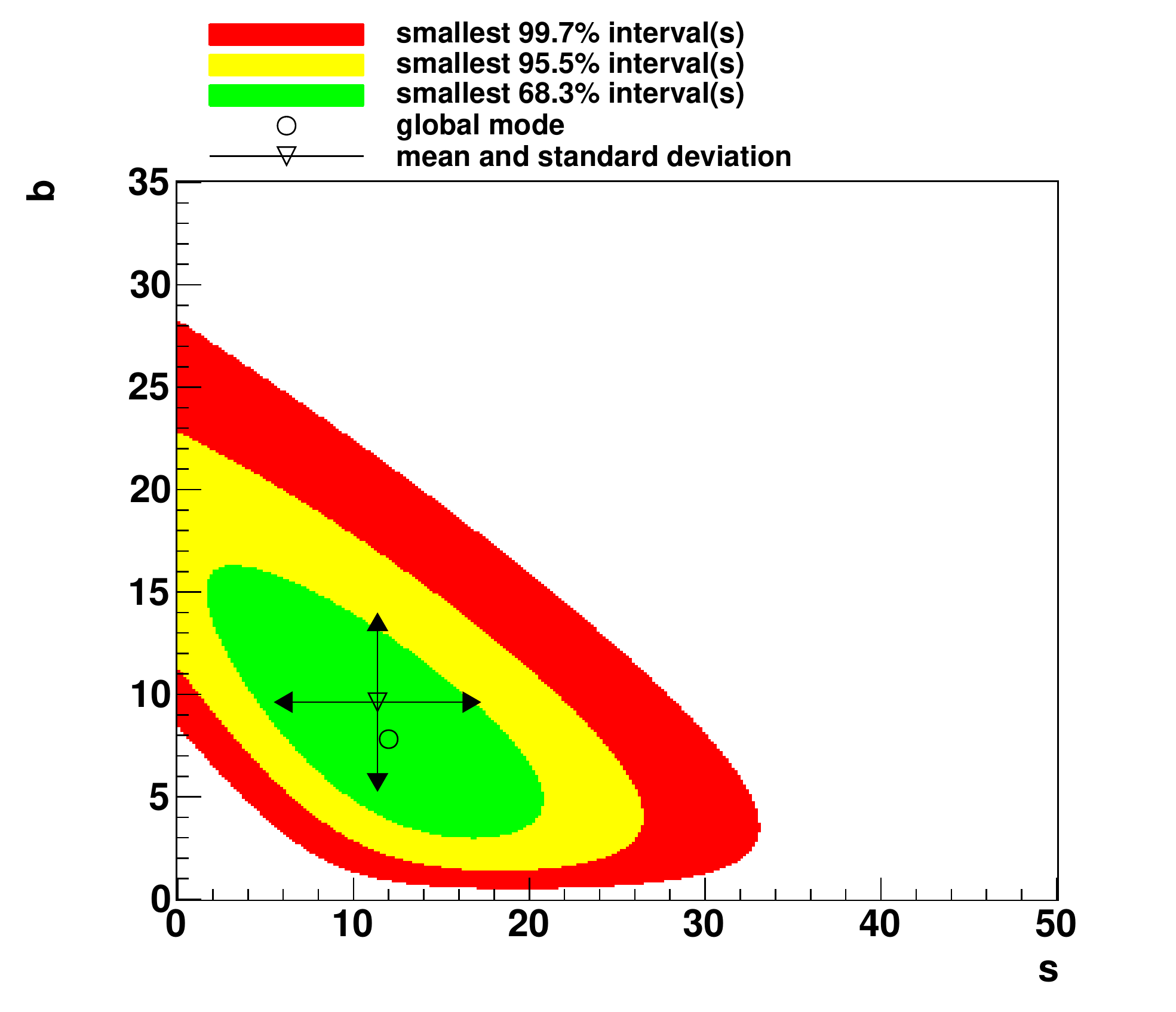}
   \end{minipage}%
   \begin{minipage}{0.5\textwidth}
     \centering
     \includegraphics[width=0.95\textwidth]{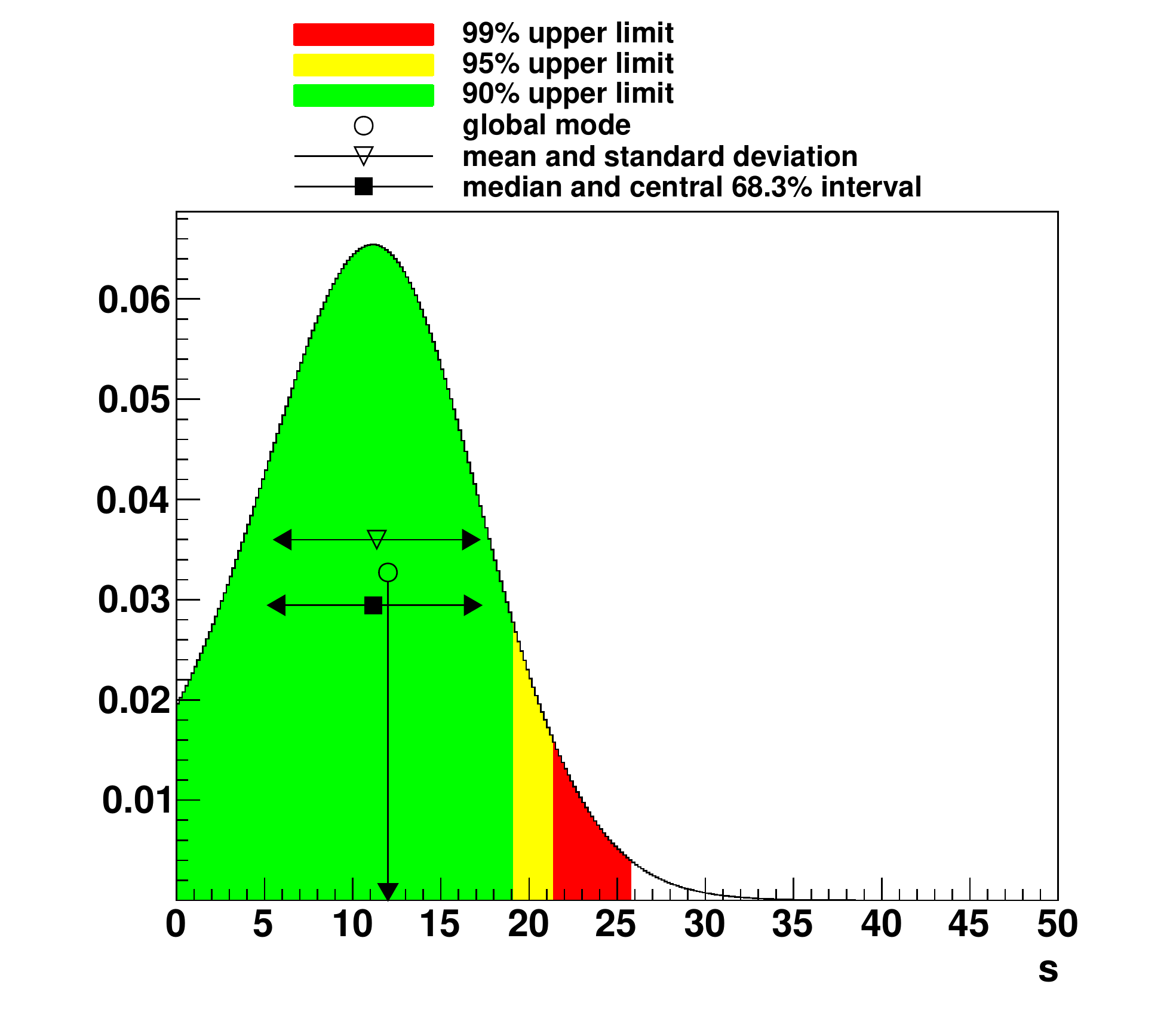}
   \end{minipage}%
   \caption{Top: prior and posterior densities of the background
     (left) and signal (right).  Bottom: joint posterior (left) and
     marginal reference posterior for the signal (right).}
   \label{fig-bat}
 \end{figure}

 After installing BAT \cite{BAT2009}, the
 \texttt{BAT/examples/advanced/referencecounting/} folder contains a
 full example.  Here we show that few C++ lines are sufficient to
 solve a typical problem (see figure~\ref{fig-bat}).

 \begin{quote}
 \footnotesize
 \begin{verbatim}
  // create new ReferenceCounting object
  ReferenceCounting* m = new ReferenceCounting();

  // BAT settings
  m->SetNbins("s", 300);
  m->SetNbins("b", 300);
  m->MCMCSetPrecision(BCIntegrate::kMedium);

  // set option of how to evaluate prior
  m->SetPriorEvalOption(ReferenceCounting::kHistogram);

  // set background 
  double bkg_exp = 10; // expectation value
  double bkg_std = 5;  // uncertainty on background
  double beta    = bkg_exp/bkg_std/bkg_std; // rate
  double alpha   = bkg_exp*beta;            // shape
  m->SetAlphaBeta(alpha, beta);

  // set number of observed events
  m->SetNObs(20);

  // set parameter range
  m->SetParameterRange(0, 0.0, 50); // signal 
  m->SetParameterRange(1, 0.0, 35); // background

  // perform sampling with MCMC
  m->MarginalizeAll();

  // perform minimization with Minuit
  m->FindMode( m->GetBestFitParameters() );

  // draw all marginalized distributions
  m->PrintAllMarginalized("plots.pdf");

  // print results of the analysis
  m->PrintResults("results.txt");
 \end{verbatim}
\end{quote}


 \section{Explicit forms for the marginal posterior}\label{sec-marg-post}

 By integrating over $b$ the left- and right-hand terms of the Bayes' theorem (\ref{eq-bayes-theorem}) one finds the marginal posterior density for $s$, proportional to the product of the marginal model (\ref{eq-marg-mod-2}) and the signal prior $p(s)$.  The result is similar to (\ref{eq-ref-posterior}).  Dropping the constant $[\beta/(1+\beta)]^\alpha$ we obtain
 \begin{equation}
   \label{eq-marginal-posterior}
   p(s|n) \propto e^{-s} \, f(s;n,\alpha,\beta) \, p(s)
 \end{equation}
 It is simple to show that the marginal model (\ref{eq-marg-mod-2})
 has the form of a linear combination of Gamma kernels.  This means
 that the posterior obtained with the uniform prior is also a mixture
 of Gamma densities.  Its form will be given below.

 Here we provide the explicit form for the marginal posterior
 (\ref{eq-marginal-posterior}) when the signal prior belongs to the
 family of Gamma densities.  When
 \[
   p(s) = \Ga(s|A,B)
 \]
 it can be shown that the marginal posterior is a linear combination
 of Gamma densities:
 \begin{equation}
   \label{eq-marginal-post-AB}
   p_{A,B}(s|n) = \frac{1}{N} \sum_{m=0}^{n} \binom{\alpha+m-1}{m}
                  \frac{\Ga(s\,|\,n-m+A,1+B)}
                       {(1+\beta)^{m} \, (n-m+A) \, \mathrm{B}(n-m+1,A)}
 \end{equation}
 where the normalization constant is
 \[
   N = \sum_{m=0}^{n} \binom{\alpha+m-1}{m}
           [(1+\beta)^{m} \, (n-m+A) \, \mathrm{B}(n-m+1,A)]^{-1}
 \]
 with the Beta function $\mathrm{B}(x,y) = \Gamma(x) \Gamma(y) /
 \Gamma(x+y)$.  This result is important for two reasons.  First, when
 there is some prior knowledge about $s$ one should use an informative
 prior, rather than the reference prior.  As for the background, it is
 best to choose a Gamma density for the signal prior, obtaining the
 posterior analytically.  Second, because the Bayes' theorem behaves
 as a linear operator, representing the prior knowledge about $s$ with
 a linear combination of Gamma densities gives a posterior which is a
 linear combination of the posteriors computed with each individual
 Gamma prior.  This means that when (\ref{eq-marginal-post-AB}) is
 used as the prior for the next experiment, one obtains the
 corresponding posterior with a simple (although tedious) algebra.

 The marginal posterior corresponding a flat prior in $s$ can be
 obtained either by direct computation from the marginal model
 (\ref{eq-marg-mod-2}), or from (\ref{eq-marginal-post-AB}) by setting
 $A=1,B=0$:
 \begin{equation}
   \label{eq-marginal-post-01}
   p_{1,0}(s|n) = \frac{1}{N'} \sum_{m=0}^{n} \binom{\alpha+m-1}{m}
                  \frac{\Ga(s\,|\,n-m+1,1)}{(1+\beta)^{m}}
 \end{equation}
 where the normalization constant is
 \[
   N' = \sum_{m=0}^{n} \binom{\alpha+m-1}{m} (1+\beta)^{-m}
 \]


 \section{Fitting the reference prior to obtain a quick
   approximation}\label{app-data}

 A text file containing values of $\pi(s)$ at discrete values of
 $s\in[0,70]$ for all values of the background shape and rate
 parameters examined in this paper is freely available on Zenodo
 (\doi{10.5281/zenodo.11896}).  The format of the file is the
 following.

 All lines contain the same number of items, separated by white
 spaces, such that the file can be considered as a 2-dimensional
 table.  The first row is special, as it provides the header for the
 table.  The first two items in the first row are the strings
 ``shape'' and ``rate'', which represent the title of the
 corresponding columns.  Next, a number of signal values follows, from
 $s=0$ to $s=70$ at equally spaced steps.  These values are the
 locations on the positive real line at which the reference prior is
 computed.

 Starting from the second line, the format is always the same.  The
 first two values are the background shape and rate parameters with
 which the reference prior $\pi(s)$ is defined for the current row.
 Next, the value $\pi(0)$ (always one) is followed by a fixed number
 of values $\pi(s_i)$, where $s_i$ is the value at position $i+2$ in
 the first row, until $\pi(70)$ is given as last value.

 The limiting form $\pi_0(s)$ only requires the knowledge of the
 background shape $\alpha$ and rate $\beta$ parameters (first two
 entries in each row, apart from the first line), as $\pi_0(s) =
 \sqrt{E[b]/(s+E[b])}$ with $E[b]=\alpha/\beta$.  To obtain a better
 approximation, the values of $\pi(s)$ can be fitted by the function
 (\ref{eq-fit-func1}), or better with (\ref{eq-fit-func2}) where the
 third parameter can be fixed at $E=0.125$ or $E=0.100$ and initial
 values for the other two parameters $A,B$ can be set to $E[b]$ and 0
 (figure~\ref{fig-AB-B} provides more precise indications if the fit
 does not converge immediately).  Of course, the user is free to try
 different functional forms.  For example, a 3-parameter function
 which well reproduces $\pi(s)$ and is very quick to compute is
 $\exp(a x^{0.25} + b x^{0.5} + c x)$.

 As the shape of $\pi(s)$ changes quite smoothly in the
 $(\log_{10}\alpha, \log_{10}\beta)$ space (as shown in
 \href{https://www.youtube.com/watch?v=vqUnRrwinHc}{https://www.youtube.com/watch?v=vqUnRrwinHc}),
 the user shall first interpolate the values of $\pi(s)$ computed with the
 points close to her background parameters (a linear interpolation in the
 log-log parameters space is enough), before fitting the resulting
 collection of $\pi(s)$ values.


%

\end{document}